\documentclass[aps,prr,reprint,superscriptaddress,amsmath,amssymb,longbibliography]{revtex4-1}
\usepackage[utf8]{inputenc}
\usepackage[normalem]{ulem}
\usepackage{amsfonts}
\usepackage{physics}
\usepackage{graphicx}
\usepackage{overpic}
\usepackage{color}
\usepackage[dvipsnames]{xcolor}
\usepackage[hidelinks]{hyperref}
\usepackage{comment}

\usepackage[T1]{fontenc}
\usepackage{enumitem}
\usepackage{booktabs}
\usepackage{float}

\usepackage{soul}

\usepackage{dsfont}
\usepackage{qcircuit}

\usepackage[caption=false]{subfig}

\graphicspath{{}}

\newcommand{\id}{\mathds{1}}
\newcommand{\Hinf}{\hat H_\text{I}}
\newcommand{\Ggate}{\mathtt{G}}
\newcommand{\CZgate}{\mathtt{CZ}}
\newcommand{\RXgate}{\mathtt{RX}}
\newcommand{\RZgate}{\mathtt{RZ}}
\newcommand{\CPgate}{\mathtt{CP}}
\newcommand{\Hgate}{\mathtt{H}}
\newcommand{\Zgate}{\mathtt{Z}}
\newcommand{\idlegate}{\mathtt{Idle}}
\newcommand{\SWAPgate}{\mathtt{SWAP}}
\newcommand{\busygate}{\mathtt{BUSY}}
\newcommand{\cirq}{\mathsf{C}}

\makeatletter

 \definecolor{boxback}{HTML}{FFF8B5}
 
 \definecolor{applegreen}{rgb}{0, 0.5, 0.0}

\newcommand{\padA}{\affiliation{Dipartimento di Fisica e Astronomia “G. Galilei” \& Padua Quantum Technologies Research Center, Università degli Studi di Padova, Italy I-35131, Padova, Italy}}
\newcommand{\padB}{\affiliation{INFN, Sezione di Padova, via Marzolo 8, I-35131, Padova}}
\newcommand{\padC}{\affiliation{Institute for Complex Quantum Systems, Ulm University, Albert-Einstein-Allee 11, 89069 Ulm, Germany}}

\usepackage{orcidlink}
\newcommand{\orciddavide}{\orcidlink{0000-0001-8219-5806}}
\newcommand{\orciddaniel}{\orcidlink{0000-0001-7658-3546}}
\newcommand{\orcidmarco}{\orcidlink{0000-0002-3215-3453}}
\newcommand{\orcidilaria}{\orcidlink{0000-0002-3806-2034}}
\newcommand{\orcidsimone}{\orcidlink{0000-0002-8882-2169}}

\definecolor{smoothred}{HTML}{C5232F}
\definecolor{mygreen}{rgb}{0,0.5,0}
\definecolor{myblue}{rgb}{0,0,0.75}
\definecolor{mymagenta}{cmyk}{0,1,0,0.12}

\newcounter{intro}

\newcounter{introa}

\newcounter{introb}

\begin{document}
%
%
\title{Quantum circuit compilation with quantum computers}
%
%
\author{Davide Rattacaso\orciddavide}\padA \padB
\author{Daniel Jaschke\orciddaniel}\padA \padB \padC
\author{Marco Ballarin\orcidmarco}\padA \padB
\author{Ilaria Siloi\orcidilaria}\padA \padB
\author{Simone Montangero\orcidsimone}\padA \padB

\date{\today}

\begin{abstract}
\noindent 
Compilation optimizes quantum algorithms performances on real-world quantum computers. 
To date, it is performed via classical optimization strategies. 
We introduce a class of quantum algorithms to perform compilation via quantum computers, 
paving the way for a quantum advantage in compilation. We demonstrate the effectiveness of this approach via Quantum and Simulated Annealing-based compilation: we successfully compile a Trotterized Hamiltonian simulation with up to $64$ qubits and $64$ time-steps and a Quantum Fourier Transform with up to $40$ qubits and $771$ time steps. We show that, for a translationally invariant circuit, the compilation results in a fidelity gain that grows extensively in the size of the input circuit, outperforming any local or quasi-local compilation approach. 
\end{abstract}

\maketitle


\section{Introduction}

Executing a quantum algorithm on a real-world quantum computer with limited resources and efficiency necessitates a compilation process. The compilation generates a circuit implementation of the algorithm that minimizes errors or runtime when executed on the specific hardware, e.g., using the native gate set~\cite{Chong2017, maronese2021quantum, schmid2023computational}.  However, even simple compilation instances have been proven NP-complete through reduction to SAT~\cite{Botea2018OnTC}, requiring approximated solutions. Numerous classical approaches have been proposed to heuristically address the compilation of quantum algorithms, both with a hardware-agnostic approach~\cite{ Saeedi_2010, Nam2018, finigan2018qubit, li2019tackling, Wille_2019, Shi_2019, Sivarajah_2020, Zhou_2020, fösel2021quantum, Peters_2022, campbell2023superstaq, Younis2021, kremer2024practical, 
PhysRevResearch.5.023146, Kikuchi2023, PRXQuantum.5.020362} and taking into account various platform-dependent error sources~\cite{9251858,Bassler2023synthesisof,schmid2023computational, jaschke2022abinitio, Tan2024compilingquantum, baker2021exploiting}.

Assuming that the development of quantum computers will follow the trajectory of classical computing, they will eventually achieve the scalability and precision necessary to compile quantum algorithms, just as classical computers successfully compile classical algorithms~\cite{alfred2007compilers}. To date, there is no strategy to compile quantum circuits with quantum computers. Motivated by the computational advantage demonstrated by quantum algorithms in solving some combinatorial optimization instances~\cite{Ebadi_2022, PhysRevX.8.031016,cain2023quantum}, we propose a general paradigm for designing quantum compilers, i.e., quantum algorithms executed on quantum devices to compile other quantum algorithms, laying the groundwork for a potential quantum speed-up in the compilation of quantum circuits.
\begin{figure}[t]
    \centering
\includegraphics[width=\linewidth]{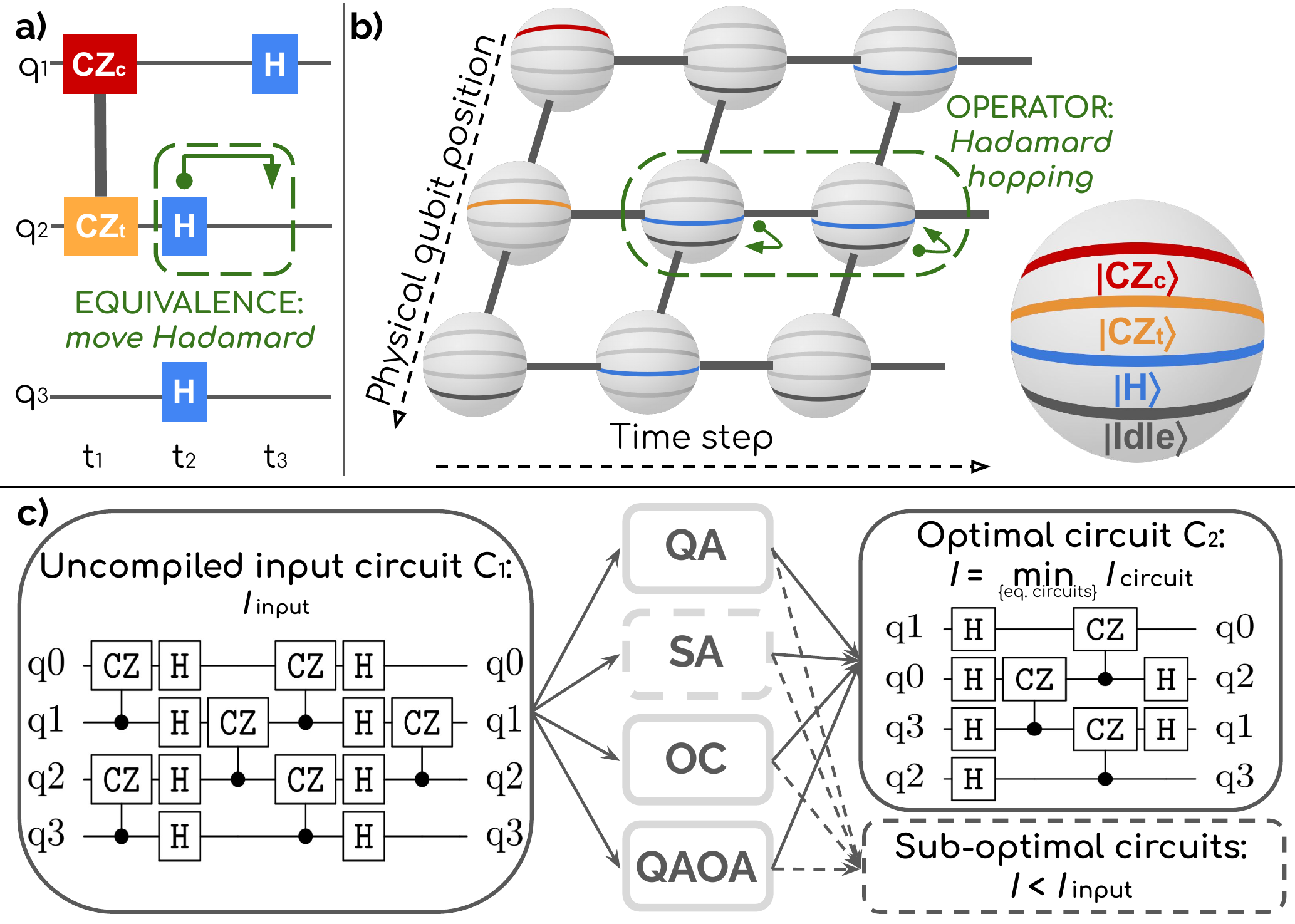}
\caption{\emph{Many-body mapping of the compilation problem.} The quantum circuit in Panel a) is mapped to the state of a lattice of qudits (spheres) in Panel b). Each internal state of a qudit represents a possible gate, e.g., Hadamard, CZ control, CZ target, or a fictitious idle gate. The local state of the qudit at site $(t,q)$ represents the gate acting at the time-step $t$ on the qubit $q$. The transformations (green dashed box) replacing equivalent sub-circuits are embedded in operators that update the states of adjacent qudits. Panel c) sketches the circuit compilation executed as a ground state search via Quantum Annealing, Simulated Annealing, Optimal Control, or QAOA. Both global optimal and sub-optimal circuits can be generated. The circuit $C_2$ is the compiled version of the circuit $C_1$, see main text.} \label{fig:mapping} 
\end{figure}
We frame the circuit compilation problem as a ground state search of a many-body \textit{infidelity Hamiltonian} $\Hinf$  diagonal in the computational basis. This Hamiltonian accounts for the errors affecting the circuit execution on the hardware --optimization task--  and enforces constraints that ensure the circuit is composed entirely by native gates --compilation task--. To this aim, each quantum circuit is encoded into a state of the computational basis of a lattice of qudits as in Figure~\ref{fig:mapping}. The ground state of 
$\Hinf$ encodes an optimal equivalent quantum circuit, representing the best arrangement of qubits and gates to minimize infidelities while producing the same quantum computation. This encoding enables the exploitation of different quantum algorithms to search for the ground state, i.e., Quantum Annealing (QA)
~\cite{Kadowaki_PRE1998,Farhi_SCI01,Santoro_SCI02,Hauke_2020,aqc_review,aqc_0}, Optimal Control (OC)~\cite{opt_c_0, opt_c_1, opt_c_2, opt_c_3, opt_c_4, Lloyd_2014, opt_c_7}, and  Quantum Approximate Optimization Algorithm (QAOA)~\cite{farhi2014quantum,PhysRevA.97.022304,farhi2019quantum,PhysRevX.10.021067,BLEKOS20241}. 

Different from standard optimization, the search space is crucially constrained to the set of many-body states representing equivalent circuits, i.e., circuits that implement the same unitary operator. To enforce this constraint during the quantum optimization, we evolve the state only through  Hamiltonians that transform circuits by replacing small sub-circuits with equivalent ones, see Figure~\ref{fig:mapping} a) and b). 
We first focus on QA and validate the QA-based quantum compiler by classically simulating the compilation of the circuit in Figure~\ref{fig:mapping} c) via tensor networks (TN)~\cite{PhysRevB.94.165116, RevModPhys.77.259, SCHOLLWOCK201196, ORUS2014117, annurev-conmatphys-040721-022705, Montangero2018, Silvi2019}. 
As it should be, the final probability of sampling the global optimal circuit increases monotonically with the annealing time (Figure~\ref{fig:quantum_annealing}). To address larger circuits, we implement Simulated Annealing (SA) based compilation, assuming its performance as a lower bound to QA. We compile quantum circuits running on a two-dimensional Quantum Processing Unit (QPU) with parametric gates and long-range crosstalk with up to $64$ qubits. The gain obtained by this compilation process scales extensively, demonstrating that this approach fully exploits the many-body nature of the compilation problem: the compilation of the global circuit is more efficient than what is obtained via combining quasi-local compiled circuits.

Beyond circuit compilation, the methods introduced in this work open up the possibility of quantum optimization of equational theories~\cite{quantum_equational_reasoning}. These formal systems automate the exploration of semantically equivalent symbolic expressions. Relying exclusively on the replacement of subexpressions, they serve as a fundamental component of symbolic algebra algorithms.

\section{Methods}

In this section, we show how the circuit compilation problem can be approached using quantum optimization techniques. In Section~\ref{subsec:infidelity}, we define a hardware-aware classical cost function that maps each candidate circuit to the expected infidelity incurred when executing it on the target hardware. In Section~\ref{subsec:circuits_as_states}, we describe how to encode quantum circuits as states of a many-qudit system. Section~\ref{subsec:inf_hamiltonian} introduces a Hamiltonian encoding of the infidelity cost function, while Section~\ref{subsec:equivalences_as_operators} presents a Hermitian operator representation of the circuit equivalence rules. Finally, in Section~\ref{subsec:compiling_as_optimization}, we demonstrate how the dynamics generated by the equivalence operators can be used to drive a quantum optimization process that minimizes the infidelity without altering the unitary operation implemented by the circuit.

\subsection{Infidelity cost function}\label{subsec:infidelity}

The objective is to find a circuit $\cirq$ that implements a target quantum algorithm while minimizing a given cost function, i.e., the infidelity, denoted as $I(\cirq)$. Each circuit $\cirq$ represents a sequence of instructions $(\Ggate, t, q)$ that specify the gate $\Ggate$ applied at the time step $t$ to the qubit $q$. We assume that the successful executions of different gates are independent events, implying that the total success probability of the circuit is the product of the individual gate fidelities~\cite{schmid2023computational}. Aiming to compile large quantum circuits, we also expect gate infidelities to be small, $i_{\Ggate} \ll 1$. Upon these assumptions, one can maximize the circuit fidelity by minimizing the sum of the infidelities affecting the native gates. Specifically, the infidelity induced by the execution of the gate $\Ggate$ on the qubit $q$ is approximated by the sum of the infidelity $i_{\Ggate}$ of the gate when executed alone, and the crosstalk contributions $x_{\Ggate}(\lVert q-q' \rVert)$ accounting for the additional infidelity that affect each couple of qubits $q$ and $q'$ involved in the simultaneous application of cross-talking gates. We also need to consider an infidelity $i_{\idlegate}$ for idle qubits, e.g., due to dephasing or decay. Under these assumptions, in Appendix~\ref{app:infidelity_derivation} we show that the total infidelity of $\cirq$ is given by:
\begin{align}\label{eq:infidelity_function}
I(\cirq)=&\sum_{(\Ggate, t, q)\in \cirq} \left(i_{\Ggate}+\sum_{q'|\Ggate(t,q')=\Ggate}x_{\Ggate}(\lVert q-q'\rVert)\right) +\\\nonumber
&\sum_{\textit{Idle }(t,q)}i_{\idlegate},
\end{align}
where $i_{\Ggate}$, $x_{\Ggate}(\lVert q-q'\rVert)$, and $i_{\idlegate}$ depend on the target hardware and are fixed parameters for the compiler.

\subsection{Encoding quantum circuits in quantum states}\label{subsec:circuits_as_states}

The number of candidate circuits to minimize infidelity grows exponentially with the number of qubits $N_Q$ and the number of time steps $N_T$ required to implement the target algorithm. We map each quantum circuit $\cirq$ to a quantum state $\ket{\cirq}$. For any QPU with physical qubits arranged on an $n$-dimensional lattice, we encode circuits in the states of the computational basis within a $(1+n)$-dimensional lattice of qudits. The first coordinate $t$ denotes the time step, while the other coordinates define a vector $\mathbf{q}$ that indicates the position of the qubit in the $n$-dimensional QPU lattice. As shown in Figure~\ref{fig:mapping}~b), each lattice site $(t, \mathbf{q})$ is then associated with a $d$-dimensional local configuration space $\{\ket{\Ggate} | \;\Ggate\in\{\idlegate,\Ggate_2, \dots, \Ggate_d\}\}$, where each state represents a gate executed at time step $t$ on qubit $\mathbf{q}$. The $\ket{\idlegate}$ state indicates that qubit $\mathbf{q}$ is idle. Circuits composed by less than $N_T$ time steps are encoded in computational basis states with some entirely idle time-steps $\ket{\idlegate}_{t,1}\otimes\dots\otimes\ket{\idlegate}_{t,N_Q}$. These idle time steps are assumed to be skipped during circuit execution. The space spanned by the computational basis representing all possible circuits is then the Hilbert space $\mathbb{H}=(\mathbb{C}^{d})^{N_Q\times N_T}$.

\subsection{The infidelity Hamiltonian}\label{subsec:inf_hamiltonian}

We define the \textit{infidelity Hamiltonian} $\Hinf$ that associates to each circuit state $\ket{\cirq}$ the infidelity $I(\cirq)=\bra{\cirq}\Hinf\ket{\cirq}$. To this aim, we introduce the operator $\hat{\mathtt{g}}_{t,q}^\dag$ that creates the gate $\Ggate$ at the lattice site $(t,q)$, whose vacuum state is the $\idlegate$ gate, and acts trivially on the other sites:
\begin{equation}
\hat{\mathtt{g}}_{t,q}^\dag := \bigotimes_{t'\neq t, q'\neq q}\id_{t',q'}\otimes\ket{\Ggate}_{t,q}\bra{\idlegate}_{t,q} \;.
\end{equation}
Any $n$-qubit gate is created (annihilated) by the action of $n$ creation (annihilation) operators. Then, $\Hinf$ yields: 
\begin{align}
\label{eq:inf_ham_quantum}
\Hinf=&\sum_{t, q}i_{\Ggate}\hat{\Ggate}_{t,q} + \sum_{t, q, \Ggate'}\sum_{q'\neq q}x_{\Ggate}(q,q')\hat{\Ggate}_{t,q}\hat{\Ggate'}_{t,q'}\\ \nonumber
&-N_Qi_\idlegate\sum_{t}\bigotimes_{q}\widehat {\idlegate}_{t,q}\;,
\end{align}
where $\hat \Ggate_{t,q} := \hat{\mathtt{g}}_{t,q}^\dag \hat{\mathtt{g}}_{t,q}$ has the role of the number operator at the lattice site $(t,q)$ for the gate $\Ggate$. The parameters $i_\Ggate$, $x_\Ggate$, and $i_\idlegate$ encode sources of infidelity specific to the target hardware. The first term in Eq.~\eqref{eq:inf_ham_quantum} is a single-body operator encoding the infidelity of each gate when executed alone, including the infidelity of idle qubits. This term also enables the suppression of non-native gates that may be present in the input circuit by assigning them a high fictitious infidelity. The second term, a two-body interaction, estimates the contribution of crosstalk. The last term is necessary to minimize the circuit depth by enforcing parallel gate execution. It maximizes the number of completely idle time steps, which can then be skipped during the execution of the circuit. More generally, we can encode various classical functions of the circuit into the Hamiltonian $\Hinf$, such as depth or energy consumption. These will remain important metrics even in the fault-tolerant era.

\subsection{Circuits equivalences as Hermitian operators}\label{subsec:equivalences_as_operators}

We leverage quantum optimization to target the equivalent circuit that minimizes the expectation value of $\Hinf$ while implementing the same unitary operator. We define a sub-circuit as any block of adjacent gates in the lattice of qudits. Starting from the original input circuit, we explore equivalent circuits by applying a set of invertible transformation rules $\mathcal{E}:=\{T=(\cirq_\text{sub}^\text{in}\leftrightarrow\cirq_\text{sub}^\text{out})\}$ that replace a sub-circuit $\cirq_\text{sub}^\text{in}$ with an equivalent one $\cirq_\text{sub}^\text{out}$ acting on the same qubits and time-steps. For example, they can include transformations that swap the execution times of commuting gates or transformations that replace non-native gates with a combination of native gates. The transformation rules used hereafter are reported in the Appendix~\ref{app:trans_rules}. Despite their validity, which can be verified through the unitary representation of the involved sub-circuits, these rules are not known to form a complete equational theory, i.e., they may not be able to generate all possible equivalent circuits. Finite and complete sets of rewriting rules have recently been established~\cite{complete_equational_theory_1,complete_equational_theory_2}.
Any finite set of rewriting rules can be incorporated into our paradigm. We create (annihilate) sub-circuits by acting with clusters of gate creation (annihilation) operators and label them via the first time-step and qubit $(t, q)$ they are acting on. The creation operator for a sub-circuit $\cirq_\text{sub}^A$ at $(t, q)$ is then:
\begin{equation}
\widehat {\cirq_\text{sub}^A(t,q)}^\dag:=\bigotimes_{t',q'}\hat {\mathtt{g}}_{t+t',q+q'}^{(A,t',q')\dag} \;,
\end{equation}
where $0\leq t'<N_T^A$, $0\leq q'<N_Q^A$, and $N_T^A$ and $N_Q^A$ are the number of time-steps and qubits included in the sub-circuit $\cirq_\text{sub}^A$. $\Ggate^{(A,t,q)}$ is the gate executed at time $t$ and qubit $q$ in $\cirq_\text{sub}^A$. 
Finally, a transformation rule $T=(\cirq_\text{sub}^\text{in}\leftrightarrow\cirq_\text{sub}^\text{out})$ corresponds to the operator
\begin{equation}
\hat T:=\sum_{\substack{t \leq N_T-N_T^\text{in},\\\, q \leq N_Q-N_Q^\text{in}}}\widehat {\cirq_\text{sub}^\text{out}(t,q)}^\dag \widehat {\cirq_\text{sub}^\text{in}(t,q)}+\text{h.c.}\;,
\end{equation}
that replaces any sub-circuit $\cirq_\text{sub}^\text{in}$ with $\cirq_\text{sub}^\text{out}$ and vice versa. Here, we assume that transformations act uniformly on the entire lattice.

\subsection{Compiling via quantum optimization}\label{subsec:compiling_as_optimization}

The optimally compiled circuit can now be singled out on a quantum computer by leveraging ground state search algorithms such as QA, QAOA, and OC. In the following, we focus on QA-based quantum compilation. The lattice of qudits is initialized in the separable state representing the input circuit. We first prepare a superposition of equivalent circuits by driving the adiabatic transition from the ground state of the single-site Hamiltonian $\hat H_0$, i.e., the input circuit, to the ground state of the driving Hamiltonian, $\hat H_\text{d} = \sum_{T\in\mathcal{E}} \hat T$. Finally, from the superposition state, we obtain the optimally compiled circuit as the ground state of the infidelity Hamiltonian $\Hinf$ by slowly turning off the driving Hamiltonian. The QA scheme reads:

\begin{equation}
  \hat H(t) = \begin{cases}
        (1-\frac{2t}{\tau}) \hat H_0 + \frac{2t}{\tau} \hat H_\text{d}  &\text{if $0\leq t \leq \frac{\tau}{2}$,}
        \\
        (2-\frac{2t}{\tau}) \hat H_\text{d} + (\frac{2t}{\tau} - 1) \Hinf &\text{if $\frac{\tau}{2}\leq t \leq \tau$\;.}
        \end{cases}
\end{equation} 
We stress that the dynamics induced by $H(t)$ is, by construction, constrained within the set of equivalent circuits. Indeed, the driving Hamiltonian $H_d(t)$ never couples states corresponding to non-equivalent circuits. Consequently, $H(t)$ is block-diagonal, with each block acting exclusively on the subspace spanned by a family of equivalent circuits.  
Since $\hat H(t)$ is a combination of local and sparse Hermitian operators, it can be efficiently simulated with a universal quantum computer~\cite{Lloyd_96, Aharonov_2003} (see Appendix~\ref{app:resurces} for details on the implementation).

\begin{figure}
    \centering
    \includegraphics[height=140pt]{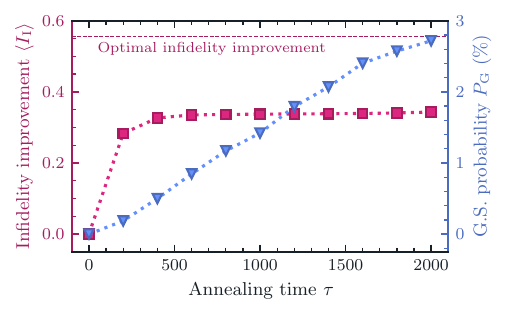}
    \caption{\emph{Quantum Annealing based compiler.} Expectation value of the infidelity improvement $I_\mathrm{I}=1-I_\mathrm{opt}/I_\mathrm{input}$ as a function of the annealing time (left y-axis). Probability $P_\mathrm{G}$ of obtaining an equivalent globally optimal circuit (right y-axis). The input circuit and the optimal circuit are depicted in Figure~\ref{fig:mapping} c). The optimal infidelity improvement refers to the optimal output circuit.}
    \label{fig:quantum_annealing}
\end{figure}

\section{Numerical results}

This section presents a numerical validation of our compilation framework. In Subsection~\ref{subsec:qa}, we simulate quantum annealing–based compilation using tensor network techniques, showing that the probability of sampling optimal equivalent circuits increases with the annealing time. In Subsection~\ref{subsec:sa}, we apply a simulated annealing version of the proposed algorithm to tackle larger compilation instances. Our results reveal the presence of local minima in the cost landscape and further motivate the use of quantum optimization techniques.

\subsection{Quantum annealing}\label{subsec:qa}

To validate this protocol in a simple setting, we consider a device consisting of a chain of four qubits with nearest-neighbor connectivity and the set of native gates  \{$\Hgate$, $\CZgate$, $\SWAPgate$\}. The 1D input circuit is depicted in Figure~\ref{fig:mapping} c) and includes six time steps. The optimization problem is encoded in an $8\times 4$ qudits lattice: the two additional steps at the beginning and the end of the computation are included to allow for swap operations. This allows for an automated search of the optimal qubit labeling. The specific gate infidelities and the local equivalence rules defining the driving Hamiltonian are listed in Appendix~\ref{app:trans_rules}, along with all details about the examples illustrated hereafter.

To transform the original input circuit into the compiled circuit shown in Figure~\ref{fig:mapping}~c), the adiabatic dynamics generated by the operators $\hat T$ and $\Hinf$ combine many quantum state transitions encoding diverse circuit transformations. Specifically, the dynamics: i) reschedules the execution time of the different gates, ii) creates and destroys pairs of $\CZgate$ and $\Hgate$ gates that are equivalent to the identity, iii) synthesizes $\SWAPgate$ gates from equivalent subcircuits composed of many $\CZgate$ and $\Hgate$ gates, and iv) moves the $\SWAPgate$ gates to the first and last time-steps. This last mechanism is enforced by a penalty term in the infidelity Hamiltonian, which assigns high infidelities to non-$\SWAPgate$ gates and zero infidelity to $\SWAPgate$ gates appearing in the first and last time steps of the circuit. Indeed, once the $\SWAPgate$ gates are moved to the external time slices, they do not need to be executed but can be interpreted as a reordering of the quantum wires of the circuit, as in Figure~\ref{fig:mapping} c). This $\SWAPgate$-based reordering then optimally associates the quantum memory addresses with the physical qubits of the machine.

We simulate the QA dynamics via a TN emulator~\cite{qtealeaves_v0_5_12, Ballarin2021, Ballarin2024}. We exploit the reflection symmetries of the problem to represent the system as a 1-dimensional lattice of $8$ qudits with local dimension $10$ interacting via up to four-body terms (see Appendix~\ref{app:example_1} for details).
In Figure~\ref{fig:quantum_annealing}, we analyze the performance of the QA based compilation. On the left y-axis, we show the \textit{infidelity improvement} $I_\mathrm{I}=1-\frac{I_\mathrm{opt}}{I_\mathrm{input}}$, where $I_\mathrm{opt}$ is the infidelity of the compiled circuit and $I_\mathrm{input}$ is the infidelity of the input circuit. As expected, $I_\mathrm{I}$ increases with the annealing time $\tau$. On the right y-axis, we plot the probability $P_\mathrm{G}$ of sampling the global optimal circuit at the end of the annealing process. This probability increases monotonically with the annealing time, reaching $3\%$ at $\tau\approx2000$. Thus, for $\tau > 500$, it is sufficient to repeat the computation $O(100)$ times to obtain the optimal solution.

\subsection{Simulated annealing}\label{subsec:sa}

Being limited by the computational complexity of emulating quantum dynamics on a classical computer, we exploit SA as an alternative strategy~\cite{doi:10.1126/science.220.4598.671}. The search for a ground state of the infidelity Hamiltonian is performed with a Markov chain Monte Carlo method (MCMC)~\cite{doi:10.1126/science.220.4598.671} where the update rules are the equivalence transformations in $\mathcal{E}$.

\begin{figure}
    \centering
    \includegraphics[height=150pt]{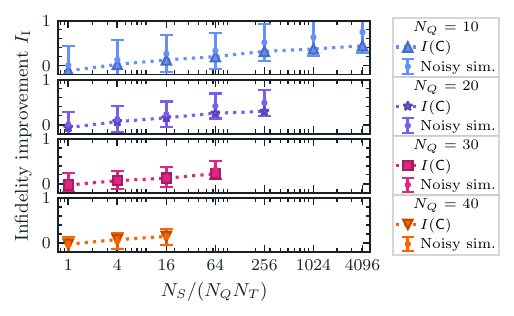} 
    \caption{
    \emph{Infidelity improvement for the QFT circuit compiled with SA.}
    Infidelity improvement $I_\mathrm{I}=1-I_\mathrm{opt}/I_\mathrm{input}$ as a function of the number of steps $N_S$ in the Markov chain over the circuit volume $N_Q N_T$. In the different Panels, circuits with different numbers of qubits $N_Q\in\{10,20,30,40\}$. We compare the infidelity improvement based on the cost function $I(C)$ with the infidelity improvement based on the noisy circuit simulation.
    }
    \label{fig:qft}
\end{figure}

\emph{Quantum Fourier Transform $-$} First, we consider the Quantum Fourier Transform (QFT)~\cite{nielsen_chuang} implemented on a quantum computer consisting of a one-dimensional qubit lattice with nearest-neighbor interactions, and a gate set composed of $\Hgate$, $\RZgate(2\pi/2^n)$, $\CZgate$, and $\SWAPgate$ gates. We assume that the main sources of errors come from $\SWAPgate$ gates and crosstalk. Figure~\ref{fig:qft} shows that SA-based compilation succeeds in reducing the expected infidelity by $65\%$, and compares the infidelity cost function $I(\cirq)$ with the infidelity affecting the execution of uncompiled and compiled circuits on a noisy machine. The latter is simulated through the noisy circuit emulator \emph{Quantum matcha TEA}~\cite{qmatchatea}. This comparison validates the correctness of $I(\cirq)$ as an estimate for the expected experimental infidelity, thus confirming that real noisy quantum computations can benefit from the proposed compilation method.

\begin{figure}
    \centering
    \includegraphics[height=150pt]{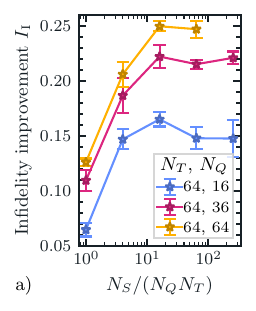}
    \hfill
    \includegraphics[height=150pt]{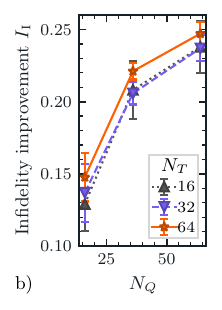}
    \caption{
    \emph{Infidelity improvement for the Trotterized Hamiltonian simulation circuit compiled with Simulated Annealing.} Panel a): infidelity improvement $I_\mathrm{I}=1-I_\mathrm{opt}/I_\mathrm{input}$ as a function of the number of steps $N_S$ in the Markov chain over the circuit volume $N_Q N_T$. We consider circuits with $N_T=64$ time-steps and different numbers of qubits $N_Q$. Panel b): infidelity improvement for the largest number of steps of the Markov chain as a function of the number of qubits $N_Q$, for different numbers of time-steps $N_T$. Results are averaged over $5$ executions of the SA.
    }
    \label{fig:ths}
\end{figure}

\emph{$2$-dimensional Trotterized Hamiltonian simulation $-$} As a second large-scale example,  we compile a Trotterized Hamiltonian simulation (THS) circuit~\cite{Lloyd_96}. THS algorithms enable the simulation of Hamiltonian evolution on digital quantum computers, providing an exponential advantage over classical simulation methods for sufficiently entangled systems~\cite{Zhou2020}. We compile a prototypical circuit that simulates the action of a suitable Hamiltonian with nearest-neighbor interactions on a two-dimensional spin lattice. The circuit is illustrated in Appendix~\ref{app:example_2}. The target quantum device also consists of a two-dimensional lattice with nearest-neighbor connectivity and parametric gates $\RZgate(\theta_1)$, $\RXgate(\theta_2)$, and $\CPgate(\theta_3)$. Parametric gates are represented in the qudit lattice encoding by discretizing the parameters $\theta_i$, and assigning a different qudit state to each possible discretized gate. The main sources of infidelity are decoherence affecting idle qubits and crosstalk between pairs of entangling gates, that decay as $\lVert \mathbf{q}-\mathbf{q}'\rVert^{-6}$ mimicking the behavior of a Rydberg atoms quantum computer~\cite{jaschke2022abinitio}. In Figure~\ref{fig:ths}, we depict the infidelity improvement for different circuit sizes and different numbers of steps in the MCMC. We observe that the improvement $I_\mathrm{I}$ at the end of the compilation process increases with the size of the input circuit. For example, $I_\mathrm{I}$ reaches about $15\%$ for a $16$-qubit circuit, while it reaches about $25\%$ for a $64$-qubit circuit. This extensive improvement can be understood by recalling that the global optimal circuit is encoded in the ground state of a translationally invariant many-body Hamiltonian. The ground state of such a system cannot be approximated by the product of the ground states of its subsystems, which is indeed a local minima. Similarly, the improvement achievable by compiling a quantum circuit is greater than that obtained by compiling its individual sub-circuits separately. 
Since quantum annealing can outperform simulated annealing in escaping local minima~\cite{Ebadi_2022,PhysRevX.8.031016,farhi2002quantumadiabaticevolutionalgorithms}, these findings provide additional motivation for the approach proposed in this work.

\section{Conclusions}

We have introduced a general paradigm for compiling quantum algorithms with quantum computers. Our approach is demonstrated with Quantum Annealing but straightforwardly extends to various techniques, including QAOA and optimal control~\cite{PhysRevA.84.012312}. We focused on optimizing equivalent circuits that implement the same unitary operator. However, as we show in Appendix~\ref{app:non_unitary}, the proposed paradigm also enables the compilation of algorithms that can be implemented by diverse unitary evolutions. For instance, quantum state preparation can be achieved by many different unitary evolutions, allowing for enlarging the circuit optimization space and potentially decreasing the optimal infidelity. Further, the presented class of compilers can provide automatic support for random circuit compilation, e.g., as needed for error mitigation~\cite{wallman2016noise}. 

Compiling a circuit requires a memory footprint that grows with the number of instructions, meaning that the compilation process demands more memory resources than the target algorithm itself. This challenge is inherent to the compilation problem and is not unique to the quantum setting. It also arises in the classical case, though it is now largely hidden by the extensive and multi-level memory architectures of modern classical devices. Hence, the compilation process will benefit from the addition of storage qubits to the QPU with multiple memory levels as in classical computers~\cite{mariantoni2011implementing, PhysRevLett.100.160501}. Meanwhile, small parts of quantum algorithms, e.g., a logical qubit in error correcting codes, can be compiled on a larger quantum machine already in the NISQ era. The compilation of larger circuits can already benefit from this many-body approach via Simulated Annealing.

Finally, the compilation paradigm demonstrated here opens a path to studying quantum algorithms as many-body systems, potentially uncovering novel phenomena like phase transitions and topological features within the space of quantum circuits. Similar phenomena have been previously demonstrated in the control parameters landscape of optimal control theory~\cite{PhysRevA.97.052114,Larocca_2018,PhysRevX.8.031086,Day_2019}.

\section*{Code and data availability}

The code implementing the algorithm for both QA and SA based circuit compilation is distributed through the VulQano \cite{vulqano_0.1.2} python package, while the code, the data to generate the
plots, and the figures of this work are available via Zenodo \cite{datasets_and_figures}.

\section*{Acknowledgments}

We acknowledge financial support from Horizon Europe programme HORIZON-CL4-2021-DIGITAL-EMERGING-01-30 via the project 101070144 (EuRyQa), the European Union via H2020 the QuantERA projects QuantHEP, T-NISQ, and QTFLAG, the Italian Ministry of University and Research (MUR) via PRIN2022 project TANQU, Fondazione CARIPARO, the INFN project QUANTUM, the Horizon 2020 research and innovation programme (Quantum Flagship - PASQuanS2), the European Union - NextGenerationEU project CN00000013 - Italian Research Center on HPC, Big Data and Quantum Computing, the Departments of Excellence grant 2023-2027 Quantum Frontiers, from the German Federal Ministry of Education and Research (BMBF) via the funding program quantum technologies - from basic research to market - project QRydDemo, and the World Class Research Infrastructure - Quantum Computing and Simulation Center (QCSC) of Padova University. We acknowledge computational resources by the Cloud Veneto, as well as computation time on Cineca’s Leonardo machine. 

\appendix

\section{Infidelity function derivation.}\label{app:infidelity_derivation}

In this appendix, we derive the infidelity function $I(\cirq)$ defined in Eq.~\eqref{eq:infidelity_function}. We assume that the probability $P(\cirq)$ that the circuit $\cirq$ returns the right output state can be written as
\begin{align}
    P(\cirq)=\prod_{\Ggate \in \cirq}F(\Ggate)\;,
\end{align}
where $F(\Ggate)$ is the fidelity of any gate $\Ggate$ in the circuit. This fidelity is eventually influenced by the parallel execution of other gates. We also associate a fidelity to idle qubits to encode, for example, the effects of decoherence.
To maximize the success probability $P(\cirq)$, we minimize the infidelity function 
\begin{align}
    I(\cirq):=-\log(P(\cirq))=-\sum_{\Ggate \in \cirq}\log(F(\Ggate))\;,
    \label{eq:logfid}
\end{align}
We introduce the gate infidelity $I(\Ggate):=1-F(\Ggate)$. For any machine capable of executing large circuits with acceptable fidelity, we can assume $I(\Ggate)\ll 1$. This implies that $\log(F(\Ggate))=\log(1-I(\Ggate))\approx -I(\Ggate)$. Replacing this approximation in the Eq.~(\ref{eq:logfid}), we obtain
\begin{align}
    I(\cirq)=\sum_{\Ggate \in \cirq}I(\Ggate)\;,
\end{align}
Finally, we write the infidelity $I(\Ggate)$ as the sum of $i_{\Ggate}$, i.e., the infidelity of the gate when executed alone, and $x_{\Ggate}(\lVert q-q' \rVert)$, accounting for the crosstalks. We also define the infidelity term $i_{\idlegate}$ for idle qubits. Replacing these equivalences in the last equation, we obtain that the infidelity function is 

\begin{align}
    I(\cirq)=\sum_{\Ggate \in \cirq} \left(i_{\Ggate}+\sum_{q,q'|\Ggate(t,q)=\Ggate(t,q')}x_{\Ggate}(\lVert q-q'\rVert)\right)\;.
\end{align}

\section{Comparative resource analysis}\label{app:resurces}

In this appendix, we compare the resources needed for the execution of the quantum algorithm with the resources required for the implementation of classical circuit compilation methods.

A fair comparison with classical strategies should take into account two main factors: (i) the computational effort needed to simulate one step of the time evolution, e.g., via Trotterized adiabatic evolution or QAOA, and (ii) the number of optimization steps needed to reach a sufficiently low expected infidelity.

The Hamiltonians involved at each time step of the evolution are: the \emph{infidelity Hamiltonian}, the \emph{driving Hamiltonian}, and the \emph{initial parent Hamiltonian}. These Hamiltonians are respectively expressed as sums of \( N_I \times V \), \( N_D \times V \), and \( N_P \times V \) operators of the form \( \ket{\alpha_1'\dots\alpha_l'}\bra{\alpha_1\dots\alpha_l} \), where \( V \) is the volume of the circuit, \( N_I \) is the number of terms per site in the infidelity Hamiltonian, \( N_D \) is the number of rewrite rules, \( N_P = 1 \) is the number of local fields used to prepare the initial state, and \( l \) is the size of a rule or a local term in the infidelity Hamiltonian.

The operators involved in a time step of QAOA or Trotterized adiabatic evolution are written as
\begin{equation}
    W = e^{-i\theta \ket{\alpha_1'\dots\alpha_l'}\bra{\alpha_1\dots\alpha_l}}.
\end{equation}
These operators can be implemented using X gates and multi-controlled NOT gates to flip an ancilla qubit when the state locally matches \( \ket{\alpha_1\dots\alpha_l} \); followed by the application of \( \mathcal{O}(l) \) single-qubit gates controlled by the ancilla to transform \( \ket{\alpha_1\dots\alpha_l} \) into \( e^{-i\theta} \ket{\alpha_1'\dots\alpha_l'} \), and finally followed by the uncomputation of the ancilla.

Multi-controlled gates can be executed efficiently on universal quantum computers using a linear number of elementary gates and ancilla qubits~\cite{PhysRevA.52.3457}, or they may be implemented natively on hardware platforms such as Rydberg-atom arrays and superconducting circuits~\cite{PhysRevX.10.021054}.

Thus, implementing each \( W \) operator requires \( \mathcal{O}(l) \) elementary gates, and a single time step of the evolution involves a total number of elementary quantum operations that scales as \( \mathcal{O}(l \cdot V) \).

On the classical side, the implementation of circuit compilation algorithms typically requires the application of equivalence rules to a data structure representing the circuit and the evaluation of the infidelity cost function. The classical computational effort required to perform these operations also increases linearly with the number of strings needed to describe the infidelity cost function, i.e., the number of operators in the infidelity Hamiltonian, and with the number of rules. Applying a transformation rule to the circuit and evaluating a cost function term are both operations whose classical cost is linear in the size of the rule or term. Finally, let us note that a similar cost is required to reduce operations on characters to Boolean logic on the classical side, and operations on qudits to standard quantum circuit operations on the quantum side. Thus, the cost of both classical or quantum operations involved in each optimization step is  $\mathcal{O}(l\cdot V)$, and the comparison of computational efforts reduces to a comparison of the number of steps needed to reach a satisfactory solution.

The number of time steps (in the quantum case) and the number of classical optimization iterations (in the classical case) needed to reach a good solution depends on the specific problem instance. This is a general feature of all quantum optimization heuristics. Demonstrating individual instances with a clear quantum speedup is a separate and high-impact challenge~\cite{Ebadi_2022,PhysRevX.8.031016,cain2023quantum}, which lies beyond the scope of this work. In this regard, our proposed compilation method—like all heuristic quantum optimization algorithms—is not designed to consistently outperform classical approaches on all instances, but rather to show a polynomial speedup in specific cases, such as when the cost function features tall, narrow energy barriers~\cite{farhi2002quantumadiabaticevolutionalgorithms} or a flat low-energy landscape~\cite{cain2023quantum}.

\section{Details of Quantum Annealing for Example I}\label{app:example_1}

In the main text, we simulate the annealing-based quantum compilation of the input circuit in Figure~1 c).
Circuits are encoded as states of an $8\times 4$ qudits lattice representing the time $t$ and the qubit $q$ as $(t,q)$. We have a local dimension of $d=5$ to represent the idle state $\idlegate$, $\Hgate$ gates, range one $\CZgate$ gates, and range one $\SWAPgate$ gates, along with a fictitious gate $\busygate$ which designates the target adjacent qubit $q+1$ for both $\CZgate$ and $\SWAPgate$. The allowed gates can be executed respectively with infidelities $i_\Hgate=0.5\cdot10^{-5}$, $i_\CZgate=1.0\cdot10^{-5}$, and $i_\SWAPgate=1.5\cdot10^{-5}$, while idle gates are affected by an infidelity $i_\idlegate=0.5\cdot10^{-5}$. The first and last time-steps are reserved for swapping area, and are associated with $0$ infidelity for $\SWAPgate$ and $\idlegate$ gates, and with infidelity $5.0\cdot10^{-5}$ to any configuration containing different gates.

With this encoding, we need to simulate the evolution in a Hilbert space whose dimension is $5^{32}$, equivalent to a $75$ qubit system. To reduce the computational complexity of the problem, we restrict the search to the space of circuits that are symmetric under the inversion of the qubit axis. This implies building symmetric $4$ qubit equivalence rules and generating the driving Hamiltonian from these transformations, which are listed in Section~\ref{app:trans_rules_1}.

Analogously, we symmetrize the infidelity Hamiltonian, while the initial Hamiltonian inherits its symmetry from the initial state. In this setting, for each time coordinate of the two-dimensional lattice, the time-evolved system state is spanned by the following ten states:
\begin{itemize}
\item $\ket{\idlegate\otimes\idlegate\otimes\idlegate\otimes\idlegate}$, \item $\ket{\Hgate\otimes\idlegate\otimes\idlegate\otimes\Hgate}$, \item $\ket{\idlegate\otimes\Hgate\otimes\Hgate\otimes\idlegate}$, \item $\ket{\Hgate\otimes\Hgate\otimes\Hgate\otimes\Hgate}$, \item $\ket{\idlegate\otimes\CZgate\otimes\busygate\otimes\idlegate}$, \item $\ket{\Hgate\otimes\CZgate\otimes\busygate\otimes\Hgate}$, \item $\ket{\idlegate\otimes\SWAPgate\otimes\busygate\otimes\idlegate}$, \item $\ket{\Hgate\otimes\SWAPgate\otimes\busygate\otimes\Hgate}$,\item $\ket{\CZgate\otimes\busygate\otimes\CZgate\otimes\busygate}$, \item $\ket{\SWAPgate\otimes\busygate\otimes\SWAPgate\otimes\busygate}$.
\end{itemize}
Note that the $\CZgate$ and $\SWAPgate$ gates are symmetric. In this way, we can represent the annealing dynamics as the evolution of a linear system of 8 qudits with a local dimension of 10, equivalent to a $\log_2(10^8)\approx27$ qubit system. The simulation time and computational complexity are driven by the sheer number of multi-qubit terms that encode the different rules; the representation of the Hamiltonian is exact and contains no truncation. To simulate the quantum evolution, we model the system state using a Matrix Product State (MPS) and evolve it through a Time-Dependent Variational Principle (TDVP) by Quantum TEA Leaves~\cite{qtealeaves_v0_5_12, PhysRevB.94.165116}, along with the collapsed representation. The effectiveness of MPS in modeling adiabatic ground state search has been previously analyzed in Ref.~\cite{Lami_2023}.

\begin{figure}
    \centering
    \includegraphics{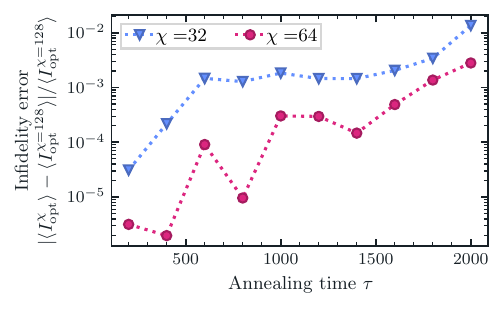}
    \caption{
    \emph{Infidelity energy convergence for different bond dimensions.} We plot the relative difference between the infidelity energy evolution evaluated with bond dimensions $\chi\in\{32,64\}$ against bond dimension $\chi=128$ as a function of the annealing time $\tau$. The number of time-steps is $N_S=4000$.
    }
    \label{fig:annealing_energy_convergence}
\end{figure}

We utilize MPS with bond dimensions $\chi\in{32, 64, 128}$ and divide the time evolution into $N_S$ time-steps with duration $\delta_t = \tau/N_S$, where $\tau$ is the annealing time and $N_S\in{1000, 2000, 4000}$. In Figure~\ref{fig:annealing_energy_convergence}, we depict the discrepancy between the simulations with bond dimensions $\chi\in\{32,64\}$ against the simulations with bond dimension $\chi=128$. For $\tau\leq 2000$, the relative error is smaller than $10^{-2}$.

\begin{figure}
    \centering    \includegraphics[width=\linewidth]{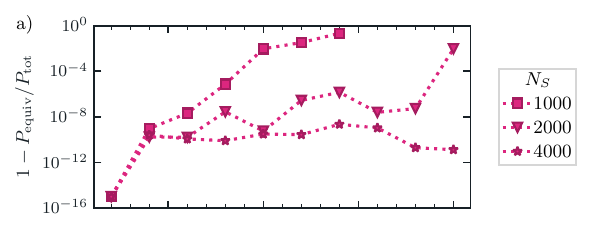}
    \hfill
    \includegraphics[width=\linewidth]{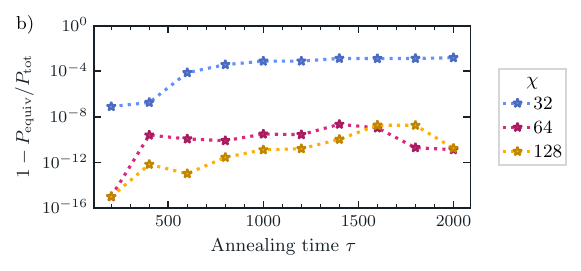} 
    \caption{
    \emph{Probability of not measuring an equivalent circuit after the Quantum Annealing-based compilation, as simulated via MPS-TDVP.} We analyze $1 - P_\text{equiv}/P_\text{tot}$, where $P_\text{equiv}$ is the sampled probability of measuring an equivalent circuit and $P_\text{tot}$ is the total sampled probability, as a function of the annealing time $\tau$. In Panel a), we fix the bond dimension $\chi=64$ and we consider different numbers of steps $N_S$. In Panel b), we fix the number of steps $N_S=4000$ and we consider different bond dimensions $\chi$.
    }
    \label{fig:annealing_convergence_dt}
\end{figure}

As a further convergence test, we perform sampling on the computational basis~\cite{Ballarin2024} of the annealed states, and we check if each sampled circuit is equivalent to the input circuit, i.e., represents the same unitary operator as the input circuit. Thus, we estimate the ratio between the sampled probability of obtaining an equivalent circuit, $P_\mathrm{equiv}$, and the total sampled probability, $P_\mathrm{tot}$. The equivalence of each sampled circuit with the input circuit is verified by checking that ${U_{\cirq_\mathrm{opt}}^\dag} U_{\cirq_\mathrm{input}}\approx\id$ up to a phase factor, where $U_{\cirq_\mathrm{input}}$ and $U_{\cirq_\mathrm{opt}}$ are the unitary operators representing the input and optimized circuits, respectively. Since the explored dynamics is constrained to the space of equivalent circuits, we expect that $1-P_\mathrm{equiv}/P_\mathrm{tot}\approx 0$ until numerical errors due to the finite number of time-steps or a small bond dimension of the MPS start accumulating. We verify the behavior of $1-P_\mathrm{equiv}/P_\mathrm{tot}$ in Figure~\ref{fig:annealing_convergence_dt}. In Panel a), we explore different numbers of steps, and in Panel b), we explore different bond dimensions. We observe that a number of steps $N_S\geq 4000$ and bond dimensions $\chi\geq 64$ are sufficient to achieve $1-P_\mathrm{equiv}/P_\mathrm{tot}<10^{-8}$, which is negligible compared to the magnitude of our main figure of merit, i.e., the probability of sampling the optimal circuit illustrated in Figure~2 of the main text.

To explain the relatively small bond dimension, note that the investigated dynamics only explores quantum superpositions of states representing circuits equivalent to the initial one. We consider a bipartition of the system into two regions $A$ and $B$, such that the time-evolved state is $\ket{\psi(t)}=\sum_{ij}M_{ij}(t)\ket{e^A_i}\otimes\ket{e^B_j}$, where ${\ket{e^A_i}}$ and ${\ket{e^B_i}}$ respectively form a computational basis for the regions $A$ and $B$. The number of non-null elements of the matrix $M_{ij}(t)$ is upper-bounded by the number $N_\text{eq}$ of equivalent circuits. The rank of $M_{ij}(t)$, which corresponds to the number of non-null singular values in the Schmidt decomposition, coincides with the number of linearly independent rows (or columns), and cannot be greater than the number of non-null elements. This constraint limits the bond dimension of the system to be upper-bounded by the number of equivalent circuits, even for long-time evolution. Tensor networks enable us to passively exploit this feature of the investigated evolution for simulation purposes.

\section{Details of Simulated Annealing for Example II}\label{app:example_2}

In the second and third example of the main text, we exploit SA to compile circuits. Beginning with the unoptimized circuit, we employ the Metropolis–Hastings algorithm to sample from the thermal states of the infidelity Hamiltonian. Here, our proposed moves for the Markov Chain are restricted solely to a set of local moves that replace equivalent sub-circuits. The temperature of the Gibbs state is gradually reduced in an effort to approach states at low temperatures, corresponding to the ground states of the infidelity Hamiltonian, and thereby representing the global optimal circuits.
In both examples, we utilize the following temperature schedule:
\begin{equation}
T_k = T_\text{max}\left(T_\text{min}/T_\text{max}\right)^{k/N}\;,
\end{equation}
where $k$ ranges from $0$ to the total number of steps of the Markov chain $N$. By increasing the total number of time steps, we effectively slow down the annealing process, resulting in an improvement of the Markov Chain's ability to escape from local minima and sample from low-temperature states.

In the second example of the main text, we focus on optimizing a circuit designed for the Trotterized simulation of a two-dimensional many-body Hamiltonian. This Hamiltonian is assumed to be time-independent and translation-invariant. The corresponding Trotterized Hamiltonian simulation circuit is represented by a state within a $(1+2)$-dimensional lattice of qudits. This state is invariant under translations of 2 sites along both qubit axis directions and under translations of 8 sites along the time-step direction. The construction details of this circuit are illustrated in Figure~\ref{fig:ths_boulding_block}. We conduct optimization procedures for Trotterized Hamiltonian simulation circuits for a QPU of $4\times 4$ and $8\times 8$ qubits. We explore various circuit depths, each representing an increasing duration of the simulated Hamiltonian evolution.

\begin{figure*}
$
\begin{array}{c}\Qcircuit @C=0.3em @R=.4em {
&\lstick{\text{q}_{2i,2j}} & \qw & \qw & \gate{\RXgate\left(\frac{\pi}{8}\right)} &\qw & \gate{\CPgate(\pi)} & \gate{\RZgate\left(\frac{\pi}{2}\right)} & \gate{\CPgate(\pi)} & \qw & \qw \\
&\lstick{\text{q}_{2i,2j+1}} & \gate{\RZgate(\pi)} & \gate{\RXgate\left(\frac{\pi}{2}\right)} & \gate{\RZgate\left(\frac{\pi}{8}\right)} & \gate{\RXgate\left(\frac{\pi}{2}\right)}& \ctrl{-1}  & \gate{\RZgate\left(\frac{\pi}{8}\right)}  & \qw & \gate{\CPgate(\pi)} & \qw \\
&\lstick{\text{q}_{2i+1,2j}} & \gate{\RZgate(\pi)} & \gate{\RXgate\left(\frac{\pi}{2}\right)} & \gate{\RZgate\left(\frac{\pi}{8}\right)} & \gate{\RXgate\left(\frac{\pi}{2}\right)}& \gate{\CPgate(\pi)} & \gate{\RZgate\left(\frac{\pi}{8}\right)} & \ctrl{-2}  & \qw & \qw \\
&\lstick{\text{q}_{2i+1,2j+1}} & \qw & \qw & \gate{\RXgate\left(\frac{\pi}{8}\right)} & \qw& \ctrl{-1} & \gate{\RZgate\left(\frac{\pi}{8}\right)} & \qw & \ctrl{-2} & \qw 
}
\end{array}
$
\caption{\emph{Fundamental building block of the Trotterized Hamiltonian simulation circuit compiled in the second example.} The circuit involves a periodic arrangement of quantum gates acting on a two-dimensional spin-lattice. The fundamental building block of this circuit is a $7$ time-steps sub-circuit operating on four adjacent qubits positioned at sites $(2i,2i+1)\times(2j,2j+1)$ within the spin lattice, where $0\leq i < r_X$ and $0\leq j < r_Y$. This building block is iteratively applied over each series of $7$ time steps from $8k$ to $8k+6$, where $0\leq k< r_T$. Additionally, $\CPgate(\pi)$ gates are applied to qubit pairs located at sites $(2i+1, j)$ and $(2i+2, j)$ at the time steps $8k+7$. These gates connect the different sub-circuits and allow for spreading entanglement among different regions of the lattice.
}
\label{fig:ths_boulding_block} 
\end{figure*}

The circuit is compiled to be optimally implemented on a quantum device consisting of a two-dimensional lattice with nearest-neighbor connectivity. We assume that the parametric gates $\RZgate(\theta)$, $\RXgate(\theta)$, and $\CPgate(\theta)$ can be executed with infidelity values of $i_\RZgate=i_\RXgate=2\cdot10^{-5}$ and $i_\CPgate=5\cdot10^{-5}$, respectively. Idle qubits are subject to decoherence, resulting in an infidelity of $i_\idlegate=1\cdot10^{-5}$ at each time step. Additionally, the simultaneous execution of a couple of $\CPgate$ gates introduces an infidelity contribution $x_\CPgate=2\cdot10^{-5}/\lVert q-q'\rVert^6$ for each pair of involved qubits, where $\lVert q-q'\rVert$ represents the lattice distance between the qubits on the device. Such small infidelities are needed to achieve satisfactory fidelity in implementing the target circuit. The local transformation rules generating the Markov Chain for the SA are listed in Section~\ref{app:trans_rules_2}.

\section{Details of the QFT for Example III}\label{app:example_3}

\begin{figure*}
\centering    \includegraphics[width=0.9\linewidth]{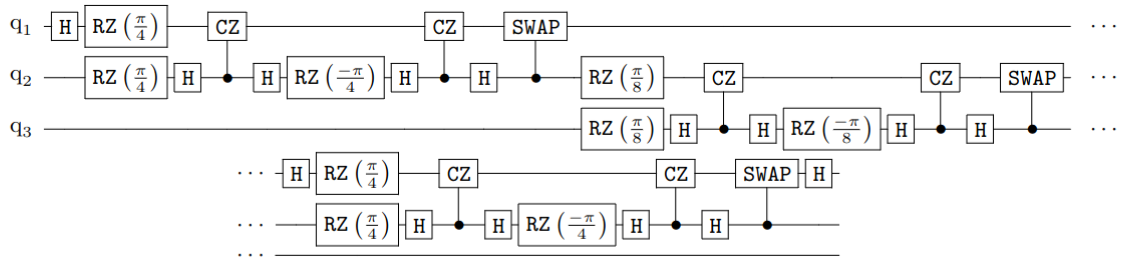}
\caption{\emph{Unoptimized $3$-qubit implementation of the QFT based on the gates $i_\Hgate$,  $i_\RZgate$,  $i_\CZgate$ and $i_\SWAPgate$.} We allow the implementation of $\SWAPgate$ gates with a relatively high infidelity, but without the crosstalk error affecting the parallel execution of $\CZgate$ gates.
}
\label{fig:qft_4} 
\end{figure*}

In the third example of the main text, we compile a Quantum Fourier Transform (QFT) implemented on a quantum computer featuring a one-dimensional qubit lattice with nearest-neighbor interactions. This device allows for the execution of $\Hgate$, $\RZgate(2\pi/2^n)$, $\CZgate$, and $\SWAPgate$ gates with the following infidelities: $i_\Hgate=1\cdot10^{-7}$, $i_\RZgate\ll 1\cdot10^{-7}$, $i_\CZgate=2\cdot10^{-7}$, and $i_\SWAPgate=20\cdot10^{-7}$. The infidelity affecting idle qubits is negligible. However, when simultaneously executing a pair of $\CZgate$ gates, an infidelity contribution $x_\CZgate=1\cdot10^{-7}/\lVert q-q'\rVert^6$ arises for each pair of involved qubits at sites $q$ and $q'$ as crosstalk. Hence, for a couple of parallel $\CZgate$ gates, $4$ crosstalk terms are taken into account. The input circuit representing the QFT is depicted in Figure~\ref{fig:qft_4} for a system of $3$ qubits, while we compile the QFT circuit for systems comprising $10$, $20$, $30$, and $40$ qubits.

Notably, to prevent the proliferation of idle gates within the circuit, which can lead to Markov chains with extensive mixing time, we introduce a fictitious \textit{lock} gate automatically positioned in the large idle regions of the input circuit. This lock gate serves as a placeholder, effectively replacing multiple idle gates. Its introduction into the circuit does not alter the cost function, but it confines the regions of the lattice where transitions of the Markov chain can occur, reducing the number of steps needed for the optimization. The local transformation rules generating the Markov Chain for the SA are listed in Section~\ref{app:trans_rules_3}.

To prove that the infidelity function $I(\cirq)$ optimized in the annealing procedure reflects the
actual fidelity of the quantum algorithm when executed on a noisy machine, we simulate the QFT circuit with noisy gates.
The simulation is carried out by Quantum Matcha TEA~\cite{qmatchatea}, a tensor network emulator
for quantum circuits, specifically using the MPS ansatz.
To simulate the noise, we employ the quantum trajectories method~\cite{trajectories_review}:  we run the circuit
for $n_{\text{traj}}$ independent times, averaging the observables over randomly distributed unitary evolutions.
We adopt a simplified error model for the gates, since we are not focused
on the physical representation of the noise at this stage, but only in its magnitude in this work. Given an $m$-qubit gate $G$, we first represent it as a parametric gate $\widetilde{G}(\theta)$ and then add Gaussian noise to the phase:
\begin{align}
    G \rightarrow \widetilde{G}(\theta+u), \quad u&\sim \mathcal{N}(0, \sigma),
\end{align}
where $\mathcal{N}(0, \sigma)$ is a normal distribution with mean $0$ and standard deviation $\sigma$. The gates are specifically handled in the following way:
\begin{itemize}
    \item We decompose the $\Hgate$ gate into:
        \begin{align}
            \Hgate& \rightarrow \mathtt{X}\mathtt{R_y}\left(\frac{\pi}{2}+u\right),\\
            \mathtt{R_y}(\theta)&=\begin{pmatrix}
            \cos(\theta/2) & -\sin(\theta/2) \\
            \sin(\theta/2) & \cos(\theta/2)
            \end{pmatrix},
        \end{align}
        where $\mathtt{X}$ is the $\mathtt{Not}$ gate.
    \item We perturb the $\RZgate$ and $\CPgate$ gates as follows:
    \begin{align}
        \RZgate(\theta) \rightarrow \RZgate(\theta+u), \quad \CPgate(\theta) \rightarrow \CPgate(\theta+u).
    \end{align}
    \item We decompose the $\SWAPgate$ gate into a circuit of $\Hgate$ and $\CZgate$ gates. We then peturb $\CZgate$ as $\CZgate\rightarrow\CPgate(\pi+u)$.
    \item We represent the crosstalk as a $4$-qubits controlled-phase gate which is close to the identity, i.e., $\mathtt{CCCP}(0+u)$.
\end{itemize}

\begin{figure}
    \centering
    \includegraphics{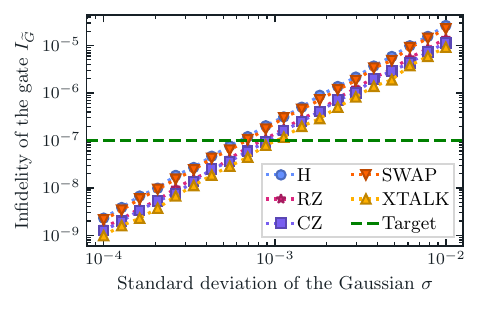}
    \caption{
    \emph{Infidelity of noisy gates with random Gaussian parameters.}
    Infidelity of the gate $I_{\widetilde{G}}$ for different standard deviations of the Gaussian
    distribution $\sigma$. We highlight with the green dashed line the target infidelity
    of $10^{-7}$.
    }
    \label{fig:gates_error}
\end{figure}

The standard deviation is chosen for each gate such that its averaged infidelity $I_{\widetilde{G}}$ is:
\begin{align}
    I(\widetilde{G}) = 1 - \frac{1}{M}\sum_{i=0}^M \left|\bra{\psi_i}G^\dag \widetilde{G}_i\ket{\psi_i}\right|^2=10^{-7},
\end{align}
where $M=1000$, $\ket{\psi_i}$ is a random state of $m$ qubits, and $\widetilde{G}_i$ represents a random sample from $\widetilde{G}$.
The infidelity of the single gate is later adjusted by the error weight of the specific gate for
that specific simulation. In Figure~\ref{fig:gates_error}, we report the assessment of the error
magnitude as a function of the standard deviation of the Gaussian random parameters.

Once this setup is defined, we can test the infidelity of a noisy quantum circuit $\cirq$, encoded by the corresponding random unitary $\widetilde{U}$,
versus the unitary $U$ representing a noiseless execution of the circuit. We obtain the infidelity of the circuit $I^\mathrm{sim}(\cirq)$ as an average
over $n_{\text{traj}}=100$ trajectories:
\begin{align}
\label{eq:simulated_infidelity}
    I^\mathrm{sim}(\cirq) = 1 - \frac{1}{n_{\text{traj}}}\sum_{i=0}^{n_{\text{traj}}} \left|\bra{\phi_i}U^\dag\widetilde{U}_i\ket{\phi_i}\right|^2,
\end{align}
where $\ket{\phi_i}$ are random separable states and $\widetilde{U}_i$ are random samples from $\widetilde{U}$.
The maximum bond dimension used in the simulation
is $\chi_{\text{max}}=128$.

\begin{figure}
    \centering
    \includegraphics{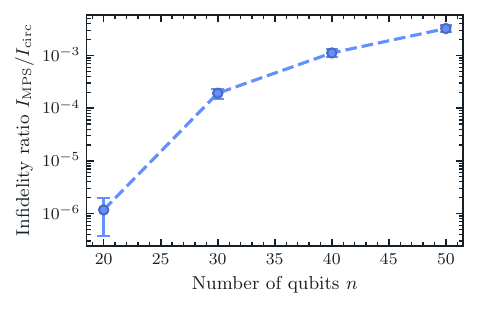}
    \caption{\emph{Infidelity from tensor network truncation error versus infidelity from noisy gates.}
    The ratio between the infidelity coming from the tensor network truncation $I_\mathrm{MPS}$ and the infidelity coming
    from the noisy gates $I_{\mathrm{circ}}$ as a function of the number of qubits. The ratio increases monotonically with the number of qubits, but it is always smaller than $10^{-2}$ for the system sizes considered here.
    }
    \label{fig:mps_errors}
\end{figure}

We stress that we select the QFT for this test since it can be efficiently simulated classically
with tensor networks~\cite{Chen2023}. However, the simulation is not trivial due to the presence
of the $4$-qubit interactions generated by the crosstalk. Thus, it is important to certify that
the infidelity we show in Figure~3 is generated by the noisy gates and not by a
tensor network truncation. By monitoring the truncation that happens in each step, i.e., in the
application of each gate, it is possible to provide a lower bound on the fidelity of the state
due to the finite bond dimension $\chi=128$.
Based on the state after $(i-1)$ multi-qubit gates, we apply the $i^{\mathrm{th}}$ multi-qubit
gate and obtain $\ket{\psi_\text{exact}^i}$ without truncation. Then, we apply our truncation
scheme and obtain $\ket{\psi_\text{trunc}^i}$. Recall that one-qubit gates do not imply additional
approximation errors; therefore, the fidelity of the $i^{\mathrm{th}}$ multi-qubit is
%
\begin{align}
F_i &=\left|\bra{\psi_\text{exact}^{i}}\ket{\psi_\text{trunc}^{i}}\right|^2=\left|\sum_{\alpha=1}^{\chi} \lambda_\alpha^2\right|^2 = \left|1-\sum_{\alpha=\chi+1}^{\chi_\text{exact}}\lambda_\alpha^2\right|^2,
\end{align}
where the $\lambda_\alpha$ are the singular values of the Schmidt decomposition of the state for the bond where the $i$-th two-qubit gate was applied, ordered by increasing magnitude, $\chi_s$ is the bond dimension of the MPS state, and $\chi_\mathrm{exact}$ is the bond dimension needed to exactly represent the state.
The infidelity of the tensor network $I_\mathrm{MPS}$ after application of the $i$-th multi-qubit gate is lower bounded as follows~\cite{jaschke2022}:
\begin{align}
\label{eq:fid_mps}
    I_\mathrm{MPS}\leq 1- \prod_{i=1}^{j-1} F_i.
\end{align}
Thus, we use this metric to ensure that the corrections due to the tensor network truncation are negligible with
respect to the errors coming from the noise. In Figure~\ref{fig:mps_errors}, we show the ratio between the infidelity
of the tensor network and the total infidelity of the simulation as defined in Equation~(3). The infidelity $I_\mathrm{MPS}$ generated by the tensor network truncation is always more than two orders of magnitude
smaller than the infidelity generated by the noise.
The point for $n=10$ qubits is not shown in Figure~\ref{fig:mps_errors}, since
the simulation is exact for this number of qubits.
We can thus state that the results we show in Figure~3 are not influenced by the finite bond dimension.

\section{Compiling more general algorithms}\label{app:non_unitary}

In the main text, we exclusively consider the compilation of quantum algorithms identifiable by a unique unitary operator, such as the Quantum Fourier Transform. However, several algorithms diverge from this framework. Examples include quantum state preparation and circuits involving ancillary qubits. For instance, the equivalence of two state-preparation algorithms hinges on their ability to prepare the same target state from a computational basis state, irrespective of their impact on other basis states. Similarly, if two circuits involve ancillary qubits, their equivalence depends solely on the output of non-ancillary qubits.

Here, we showcase the versatility of our approach by extending it to encompass these diverse quantum algorithm classes. The proposed approach is based on extending the equivalence transformations in $\mathcal{E}$ to extend the generated equivalence classes of circuits.

First, let us consider state preparation algorithms. The objective is to prepare a target state $\ket{\psi_1}$ starting from a state of the computational basis $\ket{\psi_0}$. When optimizing an input circuit that synthesizes the unitary operator $U$ to achieve $U\ket{\psi_0}=\ket{\psi_1}$, it is also essential to explore circuits that synthesize a distinct unitary $U'\neq U$ while still satisfying $U'\ket{\psi_0}=\ket{\psi_1}$. Let us illustrate this with an example involving the compilation of a quantum circuit designed to generate the GHZ state $\ket{\psi_1}=\frac{1}{\sqrt{2}}\left(\ket{0\dots0}+\ket{1\dots1}\right)$ from the computational basis state $\ket{\psi_0}=\ket{0\dots0}$. During the compilation of the input circuit, we can add or remove $\Zgate$ gates at the initial time step without altering the resulting evolution, as these gates only introduce a global phase to the initial state. Similarly, we can freely add or remove pairs of $\Zgate$ gates executed in parallel at the final time step without impacting the final state of the circuit. Thus, we expand the previously introduced transformations set $\mathcal{E}$ with additional transformations $\mathcal{E}_\text{in-sym}$ acting only on the first time-steps of the circuit. These transformations create or annihilate sub-circuits whose action does not change the input state. Similarly, we can introduce new transformations $\mathcal{E}_\text{out-sym}$ acting on the last time-steps in cases where the target state $\ket{\psi_1}$ is known. These transformations create or annihilate sub-circuits whose action leaves the target output state unchanged. By incorporating both input and output transformations in an extended set of transformation generators, we substantially expand the space of equivalent circuits. The generated circuits perform quantum state preparation through unitary transformations, expressed as $U'=\prod_{i,j} {S_\text{in}^{[i]}}^\dag U S_\text{out}^{[j]}$, where $S_\text{in}^{[i]}$ and $S_\text{out}^{[j]}$ represent unitary operators that preserve the input and target states, respectively.

When optimizing a quantum algorithm involving ancillary qubits that are not measured at the computation's conclusion, like measurement-free error correction~\cite{PRXQuantum.5.010333, veroni2024optimized}, we can enlarge the space of equivalent circuits even further. In this case, the set of transformation generators is expanded to include the subset $\mathcal{E}_\text{out-anc}$. These transformations create and annihilate arbitrary gates only in a region of the lattice corresponding to the last time steps and the ancillary qubits. This extension generates new quantum circuits that affect computational qubits similarly while altering the final state of ancillary qubits. We can also accommodate more complicated scenarios. For example, let us consider the case in which ancillary gates are prepared in a fixed state. In this situation, we can expand the set of generators to create or remove state-preserving sub-circuits at the initial time steps, specifically targeting the ancillary qubits.

In general, expanding the set of generators to accommodate a less restrictive definition of circuit equivalence enables access to new circuits that can be implemented with lower infidelity. We will explore this perspective in future work.

\section{Tranformation rules}\label{app:trans_rules}

Here, we enumerate all the transformation rules that generate the quantum and simulated annealing for the examples outlined in the main text. These rules are specific to the quantum hardware on which the circuit needs to be implemented. They encode the relationship between the native gates of the machine and those composing the input uncompiled circuit.

The complexity of the driving Hamiltonian in Quantum Annealing, in terms of the number of interactions and their range, depends on the equivalence rules and is reflected in the difficulty of simulating the Quantum Annealing dynamics using tensor networks. The number of interactions in the Hamiltonian, each corresponding to an equivalence rule applied to a region of the circuit, increases linearly with both the number of equivalence rules and the volume of the compiled circuit. Each of these interactions is an $m$-body operator, where $m$ represents the volume of the sub-circuits defining the equivalence rule.

\subsection{Example I}\label{app:trans_rules_1}

We consider the following local circuit equivalences to construct the driving Hamiltonian $\hat H_\text{d}$ for QA-based compilation of the circuit in Figure~1 c):

\begin{enumerate}
\item
$
\begin{array}{c}
\Qcircuit @C=0.5em @R=.4em {
& \gate{\Hgate} & \gate{\idlegate} & \qw \\
& \gate{\Ggate} & \gate{\Ggate} & \qw \\
& \gate{\Ggate} & \gate{\Ggate} & \qw \\
& \gate{\Hgate} & \gate{\idlegate} & \qw \\
}
\end{array} \equiv \begin{array}{c}
\Qcircuit @C=0.5em @R=.4em {
& \gate{\idlegate} & \gate{\Hgate} & \qw \\
& \gate{\Ggate} & \gate{\Ggate} & \qw \\
& \gate{\Ggate} & \gate{\Ggate} & \qw \\
& \gate{\idlegate} & \gate{\Hgate} & \qw \\
}
\end{array}
$

\item
$
\begin{array}{c}
\Qcircuit @C=0.5em @R=.4em {
& \gate{\Ggate} & \gate{\Ggate} & \qw \\
& \gate{\Hgate} & \gate{\idlegate} & \qw \\
& \gate{\Hgate} & \gate{\idlegate} & \qw \\
& \gate{\Ggate} & \gate{\Ggate} & \qw \\
}
\end{array} \equiv \begin{array}{c}
\Qcircuit @C=0.5em @R=.4em {
& \gate{\Ggate} & \gate{\Ggate} & \qw \\
& \gate{\idlegate} & \gate{\Hgate} & \qw \\
& \gate{\idlegate} & \gate{\Hgate} & \qw \\
& \gate{\Ggate} & \gate{\Ggate} & \qw \\
}
\end{array}
$

\item
$
\begin{array}{c}
\Qcircuit @C=0.5em @R=.4em {
& \gate{\Hgate} & \gate{\Hgate} & \qw \\
& \gate{\Ggate} & \gate{\Ggate} & \qw \\
& \gate{\Ggate} & \gate{\Ggate} & \qw \\
& \gate{\Hgate} & \gate{\Hgate} & \qw \\
}
\end{array} \equiv \begin{array}{c}
\Qcircuit @C=0.5em @R=.4em {
& \gate{\idlegate} & \gate{\idlegate} & \qw \\
& \gate{\Ggate} & \gate{\Ggate} & \qw \\
& \gate{\Ggate} & \gate{\Ggate} & \qw \\
& \gate{\idlegate} & \gate{\idlegate} & \qw \\
}
\end{array}
$

\item
$
\begin{array}{c}
\Qcircuit @C=0.5em @R=.4em {
& \gate{\Ggate} & \gate{\Ggate} & \qw \\
& \gate{\Hgate} & \gate{\Hgate} & \qw \\
& \gate{\Hgate} & \gate{\Hgate} & \qw \\
& \gate{\Ggate} & \gate{\Ggate} & \qw \\
}
\end{array} \equiv \begin{array}{c}
\Qcircuit @C=0.5em @R=.4em {
& \gate{\Ggate} & \gate{\Ggate} & \qw \\
& \gate{\idlegate} & \gate{\idlegate} & \qw \\
& \gate{\idlegate} & \gate{\idlegate} & \qw \\
& \gate{\Ggate} & \gate{\Ggate} & \qw \\
}
\end{array}
$

\item
$
\begin{array}{c}
\Qcircuit @C=0.5em @R=.4em {
& \gate{\Ggate} & \gate{\Ggate} & \qw \\
& \gate{\idlegate} & \gate{\CZgate} & \qw \\
& \gate{\idlegate} & \ctrl{-1} & \qw \\
& \gate{\Ggate} & \gate{\Ggate} & \qw \\
}
\end{array} \equiv \begin{array}{c}
\Qcircuit @C=0.5em @R=.4em {
& \gate{\Ggate} & \gate{\Ggate} & \qw \\
& \gate{\CZgate} & \gate{\idlegate} & \qw \\
& \ctrl{-1} & \gate{\idlegate} & \qw \\
& \gate{\Ggate} & \gate{\Ggate} & \qw \\
}
\end{array}
$

\item
$
\begin{array}{c}
\Qcircuit @C=0.5em @R=.4em {
& \gate{\CZgate} & \gate{\idlegate} & \qw \\
& \ctrl{-1} & \gate{\idlegate} & \qw \\
& \gate{\CZgate} & \gate{\idlegate} & \qw \\
& \ctrl{-1} & \gate{\idlegate} & \qw \\
}
\end{array} \equiv \begin{array}{c}
\Qcircuit @C=0.5em @R=.4em {
& \gate{\idlegate} & \gate{\CZgate} & \qw \\
& \gate{\idlegate} & \ctrl{-1} & \qw \\
& \gate{\idlegate} & \gate{\CZgate} & \qw \\
& \gate{\idlegate} & \ctrl{-1} & \qw \\
}
\end{array}
$

\item
$
\begin{array}{c}
\Qcircuit @C=0.5em @R=.4em {
& \gate{\CZgate} & \gate{\idlegate} & \qw \\
& \ctrl{-1} & \gate{\CZgate} & \qw \\
& \gate{\CZgate} & \ctrl{-1} & \qw \\
& \ctrl{-1} & \gate{\idlegate} & \qw \\
}
\end{array} \equiv \begin{array}{c}
\Qcircuit @C=0.5em @R=.4em {
& \gate{\idlegate} & \gate{\CZgate} & \qw \\
& \gate{\CZgate} & \ctrl{-1} & \qw \\
& \ctrl{-1} & \gate{\CZgate} & \qw \\
& \gate{\idlegate} & \ctrl{-1} & \qw \\
}
\end{array}
$

\item
$
\begin{array}{c}
\Qcircuit @C=0.5em @R=.4em {
& \gate{\Ggate} & \gate{\Ggate} & \qw \\
& \gate{\CZgate} & \gate{\CZgate} & \qw \\
& \ctrl{-1} & \ctrl{-1} & \qw \\
& \gate{\Ggate} & \gate{\Ggate} & \qw \\
}
\end{array} \equiv \begin{array}{c}
\Qcircuit @C=0.5em @R=.4em {
& \gate{\Ggate} & \gate{\Ggate} & \qw \\
& \gate{\idlegate} & \gate{\idlegate} & \qw \\
& \gate{\idlegate} & \gate{\idlegate} & \qw \\
& \gate{\Ggate} & \gate{\Ggate} & \qw \\
}
\end{array}
$

\item
$
\begin{array}{c}
\Qcircuit @C=0.5em @R=.4em {
& \gate{\CZgate} & \gate{\CZgate} & \qw \\
& \ctrl{-1} & \ctrl{-1} & \qw \\
& \gate{\CZgate} & \gate{\CZgate} & \qw \\
& \ctrl{-1} & \ctrl{-1} & \qw \\
}
\end{array} \equiv \begin{array}{c}
\Qcircuit @C=0.5em @R=.4em {
& \gate{\idlegate} & \gate{\idlegate} & \qw \\
& \gate{\idlegate} & \gate{\idlegate} & \qw \\
& \gate{\idlegate} & \gate{\idlegate} & \qw \\
& \gate{\idlegate} & \gate{\idlegate} & \qw \\
}
\end{array}
$

\item
$
\begin{array}{c}
\Qcircuit @C=0.5em @R=.4em {
& \gate{\CZgate} & \gate{\Hgate} & \gate{\CZgate} & \gate{\Hgate} & \qw \\
& \ctrl{-1} & \gate{\Hgate} & \ctrl{-1} & \gate{\Hgate} & \qw \\
& \gate{\CZgate} & \gate{\Hgate} & \gate{\CZgate} & \gate{\Hgate} & \qw \\
& \ctrl{-1} & \gate{\Hgate} & \ctrl{-1} & \gate{\Hgate} & \qw \\
}
\end{array} \equiv \begin{array}{c}
\Qcircuit @C=0.5em @R=.4em {
& \gate{\SWAPgate} & \gate{\Hgate} & \gate{\CZgate} & \gate{\idlegate} & \qw \\
& \ctrl{-1} & \gate{\Hgate} & \ctrl{-1} & \gate{\idlegate} & \qw \\
& \gate{\SWAPgate} & \gate{\Hgate} & \gate{\CZgate} & \gate{\idlegate} & \qw \\
& \ctrl{-1} & \gate{\Hgate} & \ctrl{-1} & \gate{\idlegate} & \qw \\
}
\end{array}
$

\item
$
\begin{array}{c}
\Qcircuit @C=0.5em @R=.4em {
& \gate{\Ggate} & \gate{\Ggate} & \gate{\Ggate} & \gate{\Ggate} & \qw \\
& \gate{\Hgate} & \gate{\CZgate} & \gate{\Hgate} & \gate{\CZgate} & \qw \\
& \gate{\Hgate} & \ctrl{-1} & \gate{\Hgate} & \ctrl{-1} & \qw \\
& \gate{\Ggate} & \gate{\Ggate} & \gate{\Ggate} & \gate{\Ggate} & \qw \\
}
\end{array} \equiv \begin{array}{c}
\Qcircuit @C=0.5em @R=.4em {
& \gate{\Ggate} & \gate{\Ggate} & \gate{\Ggate} & \gate{\Ggate} & \qw \\
& \gate{\idlegate} & \gate{\CZgate} & \gate{\Hgate} & \gate{\SWAPgate} & \qw \\
& \gate{\idlegate} & \ctrl{-1} & \gate{\Hgate} & \ctrl{-1} & \qw \\
& \gate{\Ggate} & \gate{\Ggate} & \gate{\Ggate} & \gate{\Ggate} & \qw \\
}
\end{array}
$

\end{enumerate}
The $\Ggate$ gates can be replaced with any arbitrary gate allowed by the considered architecture, such as the $H$, $\CZgate$, and $\SWAPgate$ gates, or by an $\idlegate$ qubit. This replacement must avoid any configuration that violates the qubit inversion symmetry.

\subsection{Example II}\label{app:trans_rules_2}

The local transformation rules generating the Markov chain for the SA-based compilation of THS are:
\begin{enumerate}
\item
$
\begin{array}{c}
\Qcircuit @C=0.5em @R=.4em {
&\gate{\idlegate} & \gate{\RZgate(\theta)} & \qw \\
}
\end{array}
\equiv
\begin{array}{c}
\Qcircuit @C=0.5em @R=.4em {
&\gate{\RZgate(\theta)} & \gate{\idlegate} & \qw \\
}
\end{array}
$

\item
$
\begin{array}{c}
\Qcircuit @C=0.5em @R=.4em {
& \gate{\RZgate(0)} & \qw \\
}
\end{array}
\equiv
\begin{array}{c}
\Qcircuit @C=0.5em @R=.4em {
& \gate{\idlegate} & \qw \\
}
\end{array}
$

\item
$
\begin{array}{c}
\Qcircuit @C=0.5em @R=.4em {
&\gate{\RZgate(\theta)} & \gate{\RZgate(\theta')} & \qw \\
}
\end{array}
\equiv
\begin{array}{c}
\Qcircuit @C=0.5em @R=.4em {
&\gate{\RZgate(\theta')} & \gate{\RZgate(\theta)} & \qw \\
}
\end{array}
$

\item
$
\begin{array}{c}
\Qcircuit @C=0.5em @R=.4em {
&\gate{\RZgate(\theta)} & \gate{\RZgate(\theta')} & \qw \\
}
\end{array}
\equiv
\begin{array}{c}
\Qcircuit @C=0.5em @R=.4em {
&\gate{\RZgate(\theta - \Delta)} & \gate{\RZgate(\theta' + \Delta)} & \qw \\
}
\end{array}
$

\item
$
\begin{array}{c}
\Qcircuit @C=0.5em @R=.4em {
&\gate{\idlegate} & \gate{\RXgate(\theta)} & \qw \\
}
\end{array}
\equiv
\begin{array}{c}
\Qcircuit @C=0.5em @R=.4em {
&\gate{\RXgate(\theta)} & \gate{\idlegate} & \qw \\
}
\end{array}
$

\item
$
\begin{array}{c}
\Qcircuit @C=0.5em @R=.4em {
& \gate{\RXgate(0)} & \qw \\
}
\end{array}
\equiv
\begin{array}{c}
\Qcircuit @C=0.5em @R=.4em {
& \gate{\idlegate} & \qw \\
}
\end{array}
$

\item
$
\begin{array}{c}
\Qcircuit @C=0.5em @R=.4em {
&\gate{\RXgate(\theta)} & \gate{\RXgate(\theta')} & \qw \\
}
\end{array}
\equiv
\begin{array}{c}
\Qcircuit @C=0.5em @R=.4em {
&\gate{\RXgate(\theta')} & \gate{\RXgate(\theta)} & \qw \\
}
\end{array}
$

\item
$
\begin{array}{c}
\Qcircuit @C=0.5em @R=.4em {
&\gate{\RXgate(\theta)} & \gate{\RXgate(\theta')} & \qw \\
}
\end{array}
\equiv
\begin{array}{c}
\Qcircuit @C=0.5em @R=.4em {
&\gate{\RXgate(\theta - \Delta)} & \gate{\RXgate(\theta' + \Delta)} & \qw \\
}
\end{array}
$

\item
$
\begin{array}{c}
\Qcircuit @C=0.5em @R=.4em {
&\gate{\RXgate(\theta)} & \gate{\RZgate(\pi)} & \qw \\
}
\end{array}
\equiv
\begin{array}{c}
\Qcircuit @C=0.5em @R=.4em {
&\gate{\RZgate(\pi)} & \gate{\RXgate(-\theta)} & \qw \\
}
\end{array}
$

\item
$
\begin{array}{c}
\Qcircuit @C=0.5em @R=.4em {
&\gate{\RZgate(\theta)} & \gate{\RXgate(\pi)} & \qw \\
}
\end{array}
\equiv
\begin{array}{c}
\Qcircuit @C=0.5em @R=.4em {
&\gate{\RXgate(\pi)} & \gate{\RZgate(-\theta)} & \qw \\
}
\end{array}
$

\item
\[
\begin{array}{c}
\Qcircuit @C=0.5em @R=.4em {
&\gate{\RZgate(\theta)} & \gate{\RXgate(\theta')} & \gate{\RZgate(\theta'')} & \qw \\
}
\end{array}
\equiv
\begin{array}{c}
\Qcircuit @C=0.5em @R=.4em {
&\gate{\RXgate(\gamma)} & \gate{\RZgate(\gamma')} & \gate{\RXgate(\gamma'')} & \qw \\
}
\end{array}
\]

where $(\theta,\theta',\theta'')\leftrightarrow(\gamma,\gamma',\gamma'')$ are two different Euler representations of the same rigid body rotations. The correspondence is generated by the Euler transformations between the reference frames $x-z-x$ and $z-x-z$ .

\item
$
\begin{array}{c}
\Qcircuit @C=0.5em @R=.4em {
&\gate{\idlegate} & \gate{\CPgate(\theta)} & \qw \\
&\gate{\idlegate} & \gate{\busygate} & \qw \\
}
\end{array}
\equiv
\begin{array}{c}
\Qcircuit @C=0.5em @R=.4em {
&\gate{\CPgate(\theta)} & \gate{\idlegate} & \qw \\
&\gate{\busygate} & \gate{\idlegate} & \qw \\
}
\end{array}
$

\item
$
\begin{array}{c}
\Qcircuit @C=0.5em @R=.4em {
& \gate{\CZgate(0)} & \qw \\
& \gate{\busygate} & \qw \\
}
\end{array}
\equiv
\begin{array}{c}
\Qcircuit @C=0.5em @R=.4em {
& \gate{\idlegate} & \qw \\
&\gate{\idlegate} & \qw \\
}
\end{array}
$

\item
$
\begin{array}{c}
\Qcircuit @C=0.5em @R=.4em {
&\gate{\CPgate(\theta)} & \gate{\CPgate(\theta')} & \qw \\
&\gate{\busygate} & \gate{\busygate} & \qw \\
}
\end{array}
\equiv
\begin{array}{c}
\Qcircuit @C=0.5em @R=.4em {
&\gate{\CPgate(\theta')} & \gate{\CPgate(\theta)} & \qw \\
&\gate{\busygate} & \gate{\busygate} & \qw \\
}
\end{array}
$

\item
$
\begin{array}{c}
\Qcircuit @C=0.5em @R=.4em {
&\gate{\CPgate(\theta)} & \gate{\CPgate(\theta')} & \qw \\
&\gate{\busygate} & \gate{\busygate} & \qw \\
}
\end{array}
\equiv
\begin{array}{c}
\Qcircuit @C=0.5em @R=.4em {
&\gate{\CPgate(\theta - \Delta)} & \gate{\CPgate(\theta' + \Delta)} & \qw \\
&\gate{\busygate} & \gate{\busygate} & \qw \\
}
\end{array}
$

\item
$
\begin{array}{c}
\Qcircuit @C=0.5em @R=.4em {
&\gate{\RZgate(\theta)} & \gate{\CPgate(\theta)} & \qw \\
&\gate{\idlegate} & \gate{\busygate} & \qw \\
}
\end{array}
\equiv
\begin{array}{c}
\Qcircuit @C=0.5em @R=.4em {
&\gate{\CPgate(\theta)} & \gate{\RZgate(\theta)} & \qw \\
&\gate{\busygate} & \gate{\idlegate} & \qw \\
}
\end{array}
$
\item
$
\begin{array}{c}
\Qcircuit @C=0.5em @R=.4em {
&\gate{\RZgate(\theta)} & \gate{\CPgate(\theta)} & \qw \\
&\gate{\idlegate} & \gate{\busygate} & \qw \\
}
\end{array}
\equiv
\begin{array}{c}
\Qcircuit @C=0.5em @R=.4em {
&\gate{\CPgate(\theta)} & \gate{\idlegate} & \qw \\
&\gate{\busygate} & \gate{\RZgate(\theta)} & \qw \\
}
\end{array}
$
\item
$
\begin{array}{c}
\Qcircuit @C=0.5em @R=.4em {
&\gate{\RZgate(\theta)} & \gate{\CPgate(\theta)} & \qw \\
&\gate{\RZgate(\theta)} & \gate{\busygate} & \qw \\
}
\end{array}
\equiv
\begin{array}{c}
\Qcircuit @C=0.5em @R=.4em {
&\gate{\CPgate(\theta)} & \gate{\RZgate(\theta)} & \qw \\
&\gate{\busygate} & \gate{\RZgate(\theta)} & \qw \\
}
\end{array}
$

\item
$
\begin{array}{c}
\Qcircuit @C=0.5em @R=.4em {
&\gate{\RXgate(\pi)} & \gate{\CPgate(\theta)} & \qw \\
&\gate{\idlegate} & \gate{\busygate} & \qw \\
}
\end{array}
\equiv
\begin{array}{c}
\Qcircuit @C=0.5em @R=.4em {
&\gate{\CPgate(\theta)} & \gate{\RXgate(\pi)} & \qw \\
&\gate{\busygate} & \gate{\RZgate(-\theta)} & \qw \\
}
\end{array}
$

\item
$
\begin{array}{c}
\Qcircuit @C=0.5em @R=.4em {
& \gate{\CPgate(\theta)} &\gate{\RXgate(\pi)} & \qw \\
& \gate{\busygate} & \gate{\idlegate} & \qw \\
}
\end{array}
\equiv
\begin{array}{c}
\Qcircuit @C=0.5em @R=.4em {
& \gate{\RXgate(\pi)} &\gate{\CPgate(\theta)} & \qw \\
& \gate{\RZgate(-\theta)} &\gate{\busygate} & \qw \\
}
\end{array}
$

\item
$
\begin{array}{c}
\Qcircuit @C=0.5em @R=.4em {
&\gate{\idlegate} & \gate{\CPgate(\theta)} & \qw \\
&\gate{\RXgate(\pi)} & \gate{\busygate} & \qw \\
}
\end{array}
\equiv
\begin{array}{c}
\Qcircuit @C=0.5em @R=.4em {
&\gate{\CPgate(\theta)} & \gate{\RZgate(-\theta)} & \qw \\
&\gate{\busygate} & \gate{\RXgate(\pi)} & \qw \\
}
\end{array}
$

\item
$
\begin{array}{c}
\Qcircuit @C=0.5em @R=.4em {
& \gate{\CPgate(\theta)} &\gate{\idlegate} & \qw \\
& \gate{\busygate} & \gate{\RXgate(\pi)} & \qw \\
}
\end{array}
\equiv
\begin{array}{c}
\Qcircuit @C=0.5em @R=.4em {
& \gate{\RZgate(-\theta)} &\gate{\CPgate(\theta)} & \qw \\
& \gate{\RXgate(\pi)} &\gate{\busygate} & \qw \\
}
\end{array}
$
\end{enumerate}
where angles are modulo $2\pi$, as this reflects the periodicity of the involved gates up to a global phase factor, and $\Delta\in\{\pm \pi, \pm \pi/2, \pm \pi/4, \pm \pi/8\}$.

\subsection{Example III}\label{app:trans_rules_3}

The local transformation rules generating the Markov chain for the SA-based compilation of QFT are:
\begin{enumerate}
\item
$
\begin{array}{c}
\Qcircuit @C=0.2em @R=.8em {
& \gate{\idlegate} & \gate{\Hgate} & \qw \\
}
\end{array} \equiv \begin{array}{c}
\Qcircuit @C=0.2em @R=.8em {
& \gate{\Hgate} & \gate{\idlegate} & \qw \\
}
\end{array}
$

\item
$
\begin{array}{c}
\Qcircuit @C=0.2em @R=.8em {
& \gate{\Hgate} & \gate{\Hgate} & \qw \\
}
\end{array} \equiv \begin{array}{c}
\Qcircuit @C=0.2em @R=.8em {
& \gate{\idlegate} & \gate{\idlegate} & \qw \\
}
\end{array}
$

\item
$
\begin{array}{c}
\Qcircuit @C=0.2em @R=.8em {
& \gate{\CZgate} & \gate{\CZgate} & \qw \\
& \ctrl{-1} & \ctrl{-1} & \qw \\
}
\end{array} \equiv \begin{array}{c}
\Qcircuit @C=0.2em @R=.8em {
& \gate{\idlegate} & \gate{\idlegate} & \qw \\
& \gate{\idlegate} & \gate{\idlegate} & \qw \\
}
\end{array}
$

\item
$
\begin{array}{c}
\Qcircuit @C=0.2em @R=.8em {
& \gate{\CZgate} & \gate{\idlegate} & \qw \\
& \ctrl{-1} & \gate{\CZgate} & \qw \\
& \gate{\idlegate} & \ctrl{-1} & \qw \\
}
\end{array} \equiv \begin{array}{c}
\Qcircuit @C=0.2em @R=.8em {
& \gate{\idlegate} & \gate{\CZgate} & \qw \\
& \gate{\CZgate} & \ctrl{-1} & \qw \\
& \ctrl{-1} & \gate{\idlegate} & \qw \\
}
\end{array}
$

\item
$
\begin{array}{c}
\Qcircuit @C=0.2em @R=.8em {
& \gate{\SWAPgate} & \gate{\SWAPgate} & \qw \\
& \ctrl{-1} & \ctrl{-1} & \qw \\
}
\end{array} \equiv \begin{array}{c}
\Qcircuit @C=0.2em @R=.8em {
& \gate{\idlegate} & \gate{\idlegate} & \qw \\
& \gate{\idlegate} & \gate{\idlegate} & \qw \\
}
\end{array}
$

\item
$
\begin{array}{c}
\Qcircuit @C=0.2em @R=.8em {
& \gate{\SWAPgate} & \gate{\idlegate} & \gate{\SWAPgate} & \qw \\
& \ctrl{-1} & \gate{\SWAPgate} & \ctrl{-1} & \qw \\
& \gate{\idlegate} & \ctrl{-1} & \gate{\idlegate} & \qw \\
}
\end{array} \equiv \begin{array}{c}
\Qcircuit @C=0.2em @R=.8em {
& \gate{\idlegate} & \gate{\SWAPgate} & \gate{\idlegate} & \qw \\
& \gate{\SWAPgate} & \ctrl{-1} & \gate{\SWAPgate} & \qw \\
& \ctrl{-1} & \gate{\idlegate} & \ctrl{-1} & \qw \\
}
\end{array}
$

\item
$
\begin{array}{c}
\Qcircuit @C=0.2em @R=.8em {
& \gate{\SWAPgate} & \gate{\idlegate} & \qw \\
& \ctrl{-1} & \gate{\idlegate} & \qw \\
}
\end{array} \equiv \begin{array}{c}
\Qcircuit @C=0.2em @R=.8em {
& \gate{\idlegate} & \gate{\SWAPgate} & \qw \\
& \gate{\idlegate} & \ctrl{-1} & \qw \\
}
\end{array}
$

\item
$
\begin{array}{c}
\Qcircuit @C=0.2em @R=.8em {
& \gate{\SWAPgate} & \gate{\idlegate} & \qw \\
& \ctrl{-1} & \gate{\Hgate} & \qw \\
}
\end{array} \equiv \begin{array}{c}
\Qcircuit @C=0.2em @R=.8em {
& \gate{\Hgate} & \gate{\SWAPgate} & \qw \\
& \gate{\idlegate} & \ctrl{-1} & \qw \\
}
\end{array}
$

\item
$
\begin{array}{c}
\Qcircuit @C=0.2em @R=.8em {
& \gate{\SWAPgate} & \gate{\idlegate} & \qw \\
& \ctrl{-1} & \gate{\RZgate(\pi/n)} & \qw \\
}
\end{array} \equiv \begin{array}{c}
\Qcircuit @C=0.2em @R=.8em {
& \gate{\RZgate(\pi/n)} & \gate{\SWAPgate} & \qw \\
& \gate{\idlegate} & \ctrl{-1} & \qw \\
}
\end{array}
$

\item
$
\begin{array}{c}
\Qcircuit @C=0.2em @R=.8em {
& \gate{\SWAPgate} & \gate{\Hgate} & \qw \\
& \ctrl{-1} & \gate{\idlegate} & \qw \\
}
\end{array} \equiv \begin{array}{c}
\Qcircuit @C=0.2em @R=.8em {
& \gate{\idlegate} & \gate{\SWAPgate} & \qw \\
& \gate{\Hgate} & \ctrl{-1} & \qw \\
}
\end{array}
$

\item
$
\begin{array}{c}
\Qcircuit @C=0.2em @R=.8em {
& \gate{\SWAPgate} & \gate{\Hgate} & \qw \\
& \ctrl{-1} & \gate{\Hgate} & \qw \\
}
\end{array} \equiv \begin{array}{c}
\Qcircuit @C=0.2em @R=.8em {
& \gate{\Hgate} & \gate{\SWAPgate} & \qw \\
& \gate{\Hgate} & \ctrl{-1} & \qw \\
}
\end{array}
$

\item
$
\begin{array}{c}
\Qcircuit @C=0.2em @R=.8em {
& \gate{\SWAPgate} & \gate{\Hgate} & \qw \\
& \ctrl{-1} & \gate{\RZgate(\pi/n)} & \qw \\
}
\end{array} \equiv \begin{array}{c}
\Qcircuit @C=0.2em @R=.8em {
& \gate{\RZgate(\pi/n)} & \gate{\SWAPgate} & \qw \\
& \gate{\Hgate} & \ctrl{-1} & \qw \\
}
\end{array}
$
\\
\item
$
\begin{array}{c}
\Qcircuit @C=0.2em @R=.8em {
& \gate{\SWAPgate} & \gate{\RZgate(\pi/n)} & \qw \\
& \ctrl{-1} & \gate{\idlegate} & \qw \\
}
\end{array} \equiv \begin{array}{c}
\Qcircuit @C=0.2em @R=.8em {
& \gate{\idlegate} & \gate{\SWAPgate} & \qw \\
& \gate{\RZgate(\pi/n)} & \ctrl{-1} & \qw \\
}
\end{array}
$

\item
$
\begin{array}{c}
\Qcircuit @C=0.2em @R=.8em {
& \gate{\SWAPgate} & \gate{\RZgate(\pi/n)} & \qw \\
& \ctrl{-1} & \gate{\Hgate} & \qw \\
}
\end{array} \equiv \begin{array}{c}
\Qcircuit @C=0.2em @R=.8em {
& \gate{\Hgate} & \gate{\SWAPgate} & \qw \\
& \gate{\RZgate(\pi/n)} & \ctrl{-1} & \qw \\
}
\end{array}
$

\item
$
\begin{array}{c}
\Qcircuit @C=0.2em @R=.8em {
& \gate{\SWAPgate} & \gate{\RZgate(\pi/n)} & \qw \\
& \ctrl{-1} & \gate{\RZgate(\pi/n)} & \qw \\
}
\end{array} \equiv \begin{array}{c}
\Qcircuit @C=0.2em @R=.8em {
& \gate{\RZgate(\pi/n)} & \gate{\SWAPgate} & \qw \\
& \gate{\RZgate(\pi/n)} & \ctrl{-1} & \qw \\
}
\end{array}
$

\item
$
\begin{array}{c}
\Qcircuit @C=0.2em @R=.8em {
& \gate{\CZgate} & \gate{\SWAPgate} & \qw \\
& \ctrl{-1} & \ctrl{-1} & \qw \\
}
\end{array} \equiv \begin{array}{c}
\Qcircuit @C=0.2em @R=.8em {
& \gate{\SWAPgate} & \gate{\CZgate} & \qw \\
& \ctrl{-1} & \ctrl{-1} & \qw \\
}
\end{array}
$

\item
$
\begin{array}{c}
\Qcircuit @C=0.2em @R=.8em {
& \gate{\CZgate} & \gate{\idlegate} & \gate{\SWAPgate} & \qw \\
& \ctrl{-1} & \gate{\SWAPgate} & \ctrl{-1} & \qw \\
& \gate{\idlegate} & \ctrl{-1} & \gate{\idlegate} & \qw \\
}
\end{array} \equiv \begin{array}{c}
\Qcircuit @C=0.2em @R=.8em {
& \gate{\idlegate} & \gate{\SWAPgate} & \gate{\idlegate} & \qw \\
& \gate{\SWAPgate} & \ctrl{-1} & \gate{\CZgate} & \qw \\
& \ctrl{-1} & \gate{\idlegate} & \ctrl{-1} & \qw \\
}
\end{array}
$

\item
$
\begin{array}{c}
\Qcircuit @C=0.2em @R=.8em {
& \gate{\idlegate} & \gate{\SWAPgate} & \gate{\idlegate} & \qw \\
& \gate{\CZgate} & \ctrl{-1} & \gate{\SWAPgate} & \qw \\
& \ctrl{-1} & \gate{\idlegate} & \ctrl{-1} & \qw \\
}
\end{array} \equiv \begin{array}{c}
\Qcircuit @C=0.2em @R=.8em {
& \gate{\SWAPgate} & \gate{\idlegate} & \gate{\CZgate} & \qw \\
& \ctrl{-1} & \gate{\SWAPgate} & \ctrl{-1} & \qw \\
& \gate{\idlegate} & \ctrl{-1} & \gate{\idlegate} & \qw \\
}
\end{array}
$

\item
$
\begin{array}{c}
\Qcircuit @C=0.2em @R=.8em {
& \gate{\CZgate} & \gate{\Hgate} & \gate{\CZgate} & \gate{\Hgate} & \qw \\
& \ctrl{-1} & \gate{\Hgate} & \ctrl{-1} & \gate{\Hgate} & \qw \\
}
\end{array} \equiv \begin{array}{c}
\Qcircuit @C=0.2em @R=.8em {
& \gate{\SWAPgate} & \gate{\Hgate} & \gate{\CZgate} & \gate{\idlegate} & \qw \\
& \ctrl{-1} & \gate{\Hgate} & \ctrl{-1} & \gate{\idlegate} & \qw \\
}
\end{array}
$

\item
$
\begin{array}{c}
\Qcircuit @C=0.2em @R=.8em {
& \gate{\Hgate} & \gate{\CZgate} & \gate{\Hgate} & \gate{\CZgate} & \qw \\
& \gate{\Hgate} & \ctrl{-1} & \gate{\Hgate} & \ctrl{-1} & \qw \\
}
\end{array} \equiv \begin{array}{c}
\Qcircuit @C=0.2em @R=.8em {
& \gate{\idlegate} & \gate{\CZgate} & \gate{\Hgate} & \gate{\SWAPgate} & \qw \\
& \gate{\idlegate} & \ctrl{-1} & \gate{\Hgate} & \ctrl{-1} & \qw \\
}
\end{array}
$

\end{enumerate}
$ $

\bibliography{refs_}

\end{document}